\documentstyle[eqsecnum,aps,prd,epsf]{revtex}

\begin{document}
\draft
\widetext
\title{Triad representation of the Chern-Simons state in quantum gravity}
\author{Robert Paternoga and Robert Graham}
\address{Fachbereich Physik\\ Universit\"at-GH Essen\\
45117 Essen, Germany}
\date{March 30, 2000}
\maketitle

\begin{abstract}

We investigate a triad representation of the Chern-Simons state of quantum gravity 
with a non-vanishing cosmological constant. It is shown that the Chern-Simons state, 
which is a well-known exact wavefunctional within the Ashtekar theory, 
can be transformed to the real triad representation by means of a suitably generalized
Fourier transformation, yielding 
a complex integral representation for the corresponding state
in the triad variables. It is found that topologically inequivalent choices for the
complex integration contour give rise to linearly independent
wavefunctionals in the triad representation, which all arise from the \emph{one} 
Chern-Simons state in the Ashtekar variables. 
For a suitable choice of the normalization factor, these
states turn out to be gauge-invariant under arbitrary, even topologically
non-trivial  gauge-transformations. 
Explicit analytical expressions for the wavefunctionals in the triad representation can be 
obtained in several interesting asymptotic parameter regimes, 
and the associated semiclassical 4-geometries
are discussed. In restriction to Bianchi-type homogeneous
3-metrics, we compare our results with earlier discussions 
of homogeneous cosmological models. Moreover, we define an
inner product on the Hilbert space of quantum gravity, and
choose a natural gauge-condition fixing the time-gauge.
With respect to this particular inner
product, the Chern-Simons state of quantum gravity turns out to be a 
\emph{non-normalizable} wavefunctional.

\end{abstract}




\section{Introduction}
\label{section1}

After four decades of vigorous research, a consistent quantization of general relativity
remains as one of the most fundamental problems in theoretical physics. Aside from
string theory \cite{Hen,Gre}, a promising approach to this problem is  
provided by a \emph{canonical} quantization of gravity. Since early attempts in the sixties
\cite{ADM,Wheel}, canonical quantum gravity enjoyed a renaissance after 
Ashtekar's discovery of complex spin-connection variables \cite{Ash1,Ash2},
which replaced the metric variables used so far. The new \emph{Ashekar representation} of
general relativity turned out to be closely related to a Yang-Mills theory of a local
$SO(3)$-gauge-group \cite{Ash1}, and therefore many ideas and concepts known from
Yang-Mills theory could be carried over to the theory of gravity. In particular, the 
\emph{loop representation}, which had just been investigated within Yang-Mills theory
\cite{Gam}, furnished yet another representation of general relativity
\cite{Ash1,Rov1,Thie}, and, moreover, a remarkable connection between gravity and knot 
theory \cite{Rov1,Bm1}. Later on, the loop representation of general relativity advanced to
a mathematically rigorous theory within the framework of discretized models of gravity, the
so-called quantum spin-networks \cite{Maj,Rov2}.  

As one crucial advantage of the Ashtekar representation the constraint operators of 
quantum gravity took a polynomial form in the new spin-connection variables, and 
explicit solutions were found. Among the different quantum states discussed so far
\cite{Grieg,Bm2}, the \emph{Chern-Simons state} \cite{Ble,Kod} played an outstanding 
role, since it was the only wavefunctional with a well-defined semiclassical limit.\footnote{
Strictly speaking, this is only true for a non-vanishing cosmological constant, where
deSitter-like 4-geometries are described by the semiclassical Chern-Simons state
\cite{Bm2,Smo}. The case of a vanishing cosmological constant has been investigated
by Ezawa in  \cite{Eza}, where it turned out, that the semiclassical 4-geometries will in
general suffer from different pathologies.}
A loop representation of the Chern-Simons state was investigated, and turned out to
be closely related to the Kauffman-bracket \cite{Kauf}. Moreover, this
particular state was found to make an obvious
connection between quantum gravity and topological field theory \cite{Kauf,Wit1}.

However, a physical interpretation of the Chern-Simons state within the Ashtekar
representation implied several problems, which arose from the \emph{reality conditions}
underlying Ashtekar's complex theory of gravity \cite{Ash1}. Different \emph{real}
versions of Ashtekar's theory were suggested \cite{Loll,Barb,Imm}, but the corresponding
quantum constraint equations turned out to be non-polynomial, lacking the Chern-Simons
state as a solution.  

Amazingly, a rather natural way to circumvent the problems associated with 
Ashtekar's reality conditions has never been investigated: If we would be able to
transform the Chern-Simons state from the Ashtekar to the metric representation, the
geometrical meaning of the fundamental variables would be obvious, and no
further reality conditions would be needed. In addition, also
questions concerning the normalizability of the Chern-Simons state are much easier
to discuss in the real metric variables, than in the complex Ashtekar spin-connection
variables. It is therefore interesting to find an explicit transformation connecting these two
representations, and to study the Chern-Simons state in the metric representation.

Recently, we examined this problem in the framework of the \emph{homogeneous} 
Bianchi-type~IX
model \cite{Pat1,Pat2,Pat3}. As an intermediate step, we introduced the 
\emph{triad representation} of general relativity, which is trivially connected to the
metric representation we were interested in. Then it turned out that the Chern-Simons
state in the Ashtekar representation can be transformed to the triad variables by
a suitably generalized Fourier transformation. Topologically inequivalent choices
for the \emph{complex} integration contour in the Fourier integral gave rise to different, 
linearly independent quantum states in the triad representation, which all arose from 
the \emph{one} Chern-Simons state in the Ashtekar variables. We found explicit integral
representations for the corresponding states in the triad variables, and
gave semiclassical interpretations of the wavefunctions in different asymptotic 
parameter regimes.

In the present paper, we now want to push these results for the homogeneous model
a big step further, and will ask for the corresponding form of the \emph{inhomogeneous}
Chern-Simons state in the triad representation. For technical reasons, we will restrict
ourselves to model Universes, where the spatial hypersurfaces of constant time are
compact and without boundaries, but of arbitrary topology. In order to recover the
Chern-Simons state as a quantum state of gravity, we should allow for a 
non-vanishing cosmological constant, which, by the way, is in complete agreement with
current cosmological data \cite{Perl1,Perl2}.

The rest of this paper is organized as follows: In section~\ref{section2} we define
our notation and start from the metric representation of classical general relativity. We
introduce the triad and the Ashtekar variables, and give new representations of
the constraint obser\-vables in terms of a single tensor density, which is closely related
to the curvature of the Ashtekar spin-connection. A canonical quanization of the
theory is performed in section~\ref{section3}. 
Choosing a particular factor ordering for
the constraint operators of quantum gravity, we discuss the corresponding
operator algebra, and show that it closes without any quantum corrections. The 
transformation connecting the Ashtekar and the triad representation is explained in detail,
and is then used to derive a functional integral representation for the Chern-Simons state
in the triad representation. In section~\ref{section4} we study several asymptotic expansions
of this functional integral in some physically interesting parameter regimes. In particular,
we are interested in the semiclassical form of the Chern-Simons state, which then will 
allow for a discussion of the semiclassical 4-geometries. A separate 
subsection~\ref{section4.2.0} is
dedicated to the behavior of the Chern-Simons state under large,
topologically non-trivial $SO(3)$-gauge-transformations. The value of the
Chern-Simons state on Bianchi-type
homogeneous 3-manifolds is computed and compared with earlier results obtained
within the framework of homogeneous models. In section~\ref{section5}
we define an inner product on the Hilbert space of quantum gravity, which
is gauge-fixed with respect to the time-redefinition-invariance, and examine 
the normalizability of the Chern-Simons state. Finally, 
we summarize our conclusions in section~\ref{section6}. 
Three appendices deal with certain technical
details. In appendix~\ref{appA}, we discuss the solvability of the 
saddle-point equations, which determine the semiclassical Chern-Simons state, and
show how the solutions of these equations correspond to \emph{divergence-free}
triads in the limit of a vanishing cosmological constant. In appendix~\ref{appB}, then 
five divergence-free triads are calculated for homogeneous Bianchi-type~IX metrics,
and the corresonding values of the Chern-Simons state are given. In order to comment
on possible boundary conditions satisfied by the Chern-Simons state, a further 
appendix~\ref{appC}
deals with the asymptotic behavior of particular semiclassical 4-geometries, 
which arise for a special class of initial 3-metrics.


\section{Triad representation and Ashtekar variables}
\label{section2}

In order to set the stage and to define our notation let us briefly recall the ADM Hamiltonian
formulation of general relativity \cite{ADM,MTW,KolbTur} in terms of the densitized inverse
triad ${\tilde{e}}^i \, \!_a$ and its canonically conjugate momentum
$p_{ia}$. This will be called the triad representation for short
\cite{Ash1,Kod,Barb,Ash3,Mat}.

The most commonly used form of the ADM formulation \cite{ADM} employs as
generalized coordinates the metric tensor $h_{i j}$ on a family of
space-like 3-manifolds foliating space-time. Alternatively one
may also employ the inverse metric tensor $h^{i j}$ with
$h^{i j} h_{j k}=\delta^i_k$, or, what will be done here, the
densitized inverse metric
\begin{equation}
\tilde{\tilde{a}}^{i j}=h \, h^{i j}
\label{eq:2.1}
\end{equation}
with $h=\det(h_{i j})$.\footnote{Here and in the following densities of 
positive weight are denoted by an upper, and densities of negative weight by a lower tilde.}
Then the canonically conjugate
momenta $\, \mbox{\raisebox{-1.8 ex}{$\tilde{}$} \hspace{-1.3 ex}} \pi_{i j}$, 
which form a tensor density of weight $-1$, become
\begin{equation}
 \, \mbox{\raisebox{-1.8 ex}{$\tilde{}$} \hspace{-1.3 ex}} \pi_{i j}
 =
 \frac{\delta L}{\delta \dot{\tilde{\tilde{a}}} \,\!^{^{^{i j}}}}
 =
 \frac{1}{\gamma \sqrt{h}} \, K_{i j} \ ,
\label{eq:2.2}
\end{equation}
where $\gamma=16 \pi G$ is a convenient abbreviation containing Newton's
constant $G$, and $K_{i j}$ is the usual extrinsic curvature describing the
embedding of the 3-manifold in space-time. The quantity $L$ in (\ref{eq:2.2}) is the
Lagrangian defined by the Einstein-Hilbert action \cite{MTW,KolbTur}, in which we include
a cosmological term with a cosmological constant $\Lambda$. This choice of
variables implies a symplectic structure on phase-space defined by the
Poisson-brackets
\begin{eqnarray}
 &\Bigl \{
 \tilde{\tilde{a}}^{i j}(x) \, ,\, \, \mbox{\raisebox{-1.8 ex}{$\tilde{}$} \hspace{-1.3 ex}} 
 \pi_{k \ell}(y)
 \Bigr \}
 = 
 \textstyle \frac{1}{2}
 \Bigl (
 \delta^i_k \, \delta^j_\ell + \delta^i_\ell \, \delta^j_k
 \Bigr) \delta^3 (x-y) \ , &
 \nonumber\\[.2 cm]
 &\Big \{
 \tilde{\tilde{a}}^{i j}(x) \, , \, \tilde{\tilde{a}}^{k \ell}(y)
 \Bigr \}
 = 0 =
 \Bigl \{
 \, \mbox{\raisebox{-1.8 ex}{$\tilde{}$} \hspace{-1.3 ex}} \pi_{i j}(x) \, ,
 \, \, \mbox{\raisebox{-1.8 ex}{$\tilde{}$} \hspace{-1.3 ex}} \pi_{k \ell}(y)
 \Bigr \} \ . &
 \label{eq:2.3}
\end{eqnarray}
Indices $i, j$ will be raised and lowered by $h^{i j}$ and its inverse. In
order to move on to the triad representation let us introduce the densitized
inverse triad ${\tilde{e}}^i \,\!_a$ via
\begin{equation}
 {\tilde{e}}^i \,\!_a \cdot {\tilde{e}}^j \,\!_a= \tilde{\tilde{a}}^{i j} \ ,
\label{eq:2.4}
\end{equation}
and define an enlarged phase-space by introducing canonically conjugate
momenta $p_{i a}$ of the ${\tilde{e}}^i \,\!_a$ with Poisson-brackets
\begin{eqnarray}
&\Bigl \{{\tilde{e}}^i \,\!_a (x)\, , \, p_{j b}(y) \Bigr \}
=
\delta_j^i \, \delta_{a b} \, \delta^3(x-y) \ ,& \nonumber \\[.2 cm]
&\Bigl \{ p_{i a}(x) \, , \, p_{j b}(y) \Bigr \}
=
0 \ . &
\label{eq:2.5}
\end{eqnarray}
In the following we shall also make use of the triad 1-forms $e_{i a}$ and 
the triad vectors
${e}^i \,\!_a={\tilde{e}}^i \,\!_{a} / \sqrt{h}$. In order to guarantee that
(\ref{eq:2.3}) is compatible with (\ref{eq:2.4}), (\ref{eq:2.5}), we relate
$\, \mbox{\raisebox{-1.8 ex}{$\tilde{}$} \hspace{-1.3 ex}} \pi_{i j}$ to $p_{j a}$ via
\begin{equation}
 \, \mbox{\raisebox{-1.8 ex}{$\tilde{}$} \hspace{-1.3 ex}} \pi_{i j}
 =\frac{1}{2 \sqrt{h}} \, e_{i a} \, p_{j a} \ ,
\label{eq:2.6}
\end{equation}
which serves to satisfy the first of eqs.~(\ref{eq:2.3}). Furthermore we
introduce the three additional constraints
\begin{equation}
\tilde{\cal{J}}_a :=\varepsilon_{a b c} \, {\tilde{e}}^i \,\!_b \, p_{i c}  \stackrel{!}{=} 0 \ .
\label{eq:2.7}
\end{equation}
Here the Levi-Cevitta tensor $\varepsilon_{a b c}$ is defined by
\begin{equation}
  \label{eq:2.7+}
  \varepsilon_{a b c} := \varepsilon ( e_{i a} ) \cdot [a b c] \ ,
\end{equation}
where $\varepsilon ( e_{i a} ) \in \{ \pm 1 \}$ measures the orientation of the
triad $e_{i a}$, and $[a b c]$ is the totally antisymmetric Levi-Cevitta
symbol normalized such that $[1 2 3] = +1$.
On the constraint hypersurface defined by (\ref{eq:2.7}) the quantity
$\, \mbox{\raisebox{-1.8 ex}{$\tilde{}$} \hspace{-1.3 ex}} \pi_{i j}$
defined by (\ref{eq:2.6}) is easily checked to be symmetric in $i$, $j$ as
required by (\ref{eq:2.2}) and to satisfy the last of eqs.~(\ref{eq:2.3}).

The ADM-Hamiltonian \cite{MTW,KolbTur}
\begin{equation}
H^{{\rm ADM}} = \int d^3 x
\left(N \tilde{{\cal H}}_0^{{\rm ADM}}+N^i \tilde{{\cal H}}_i \right)
\label{eq:2.8}
\end{equation}
with Lagrangian parameters $N$, $N^i$ and constraints
$\tilde{{\cal H}}_0^{{\rm ADM}}$, $\tilde{{\cal H}}_i$ given in terms of
$\tilde{\tilde{a}}^{i j}$, $\, \mbox{\raisebox{-1.8 ex}{$\tilde{}$} \hspace{-1.3 ex}} \pi_{i j}$ 
is easily rewritten in terms
of the triad representation using eqs.~(\ref{eq:2.4}), (\ref{eq:2.6}). This
yields (cf.~\cite{Ash1,Kod,Barb,Mat})
\begin{displaymath}
\tilde{{\cal H}}_0^{{\rm ADM}} 
=
-\frac{\gamma}{4} \, e_{i a} \, \tilde{\varepsilon}^{i j k} \varepsilon_{abc} \, p_{j b} \, p_{k c}+
 \frac{1}{\gamma} \, e_{i a} \, \tilde{\varepsilon}^{i j k} \, {F}_{ j k a}+
  \frac{2 \Lambda}{\gamma} \,  \sqrt{h} \ , 
\end{displaymath}
\begin{equation}
\tilde{{\cal H}}_i
=
\partial_j \left ( {\tilde{e}}^j \,\!_a \, p_{i a} \right )-
 {\tilde{e}}^j \,\!_a \, \partial_i p_{j a} \ ,
\label{eq:2.9}
\end{equation}
where $\tilde{\varepsilon}^{i j k}$ is the spatial Levi-Cevitta tensor density,\footnote{
With our definition of $\varepsilon_{a b c}$ in (\ref{eq:2.7+}) the spatial
Levi-Cevitta tensor density is naturally obtained as  
$\tilde{\varepsilon}^{i j k}=\sqrt{h} \, \varepsilon_{a b c} \, e^i \,\!_a  \, e^j \,\!_b  \, e^k \,\!_c$.}
and
${F}_{j k a}=\partial_j \omega_{k a}-\partial_k \omega_{j a}+\varepsilon_{a b c}
\omega_{j b} \omega_{k c}$ is the curvature of the Riemannian spin-connection
$\omega_{i a}=-\frac{1}{2} \, \varepsilon_{a b c} \, e_{j b} \nabla_{\! i} \, {e^j} \,\!_c$. The
additional constraints (\ref{eq:2.7}) must of course be added to the
Hamiltonian (\ref{eq:2.8}) with new Lagrangian parameters $\Omega_a$.

The introduction of the complex Ashtekar variables \cite{Ash1,Ash2,Ash4}
\begin{equation}
{\cal A}_{i a}=\omega_{i a} \pm \frac{i \gamma}{2} \, p_{i a}
\label{eq:2.11}
\end{equation}
instead of the canonical momenta $p_{i a}$ is now convenient in order to
simplify the constraints. In the framework of this paper we shall use the variables 
${\cal A}_{i a}$ just as
auxiliary quantities. In eq.~(\ref{eq:2.11}) either ``+'' or ``-''
may be chosen, but we will keep this option open by using both signs
together. The two choices are classically equivalent, but lead to inequivalent
quantizations in the quantum theory. The Poisson-brackets in the new
variables then take the form
\begin{equation}
\label{eq:2.18a}
\Bigl \{ {\tilde{e}}^i \,\!_a(x)\, , \, {\cal A}_{j b}(y) \Bigr \}
=
\pm \frac{i \gamma}{2} \, \delta_j^i \, \delta_{a b} \, \delta^3(x-y) \ , 
\end{equation}
\begin{equation}
\label{eq:2.18b}
\Bigl \{ {\cal A}_{i a}(x) \, ,\, {\cal A}_{j b}(y) \Bigr \}
= 0 \ .
\end{equation}
The second of these relations follows from the fact that the Riemannian
spin-connection $\omega_{i a}$ can be expressed as \cite{Kod}
\begin{equation}
\omega_{i a}=\frac{\delta \phi}{\delta {\tilde{e}}^i \,\!_a}
\label{eq:2.19a}
\end{equation}
with
\begin{equation}
\phi :=-{\textstyle \frac{1}{2}} \int d^3 x \, \tilde{\varepsilon}^{i j k}  e_{i a} \, \partial_j e_{k a}\ .
\label{eq:2.20a}
\end{equation}
Employing ${\cal A}_{i a}$ as a new and complex spin-connection it is convenient
to use also its associated curvature
\begin{equation}
{\cal F}_{i j a}=\partial_i {\cal A}_{j a}-\partial_j {\cal A}_{i a}+ \varepsilon_{a b c} \,
{\cal A}_{i b} \, {\cal A}_{j c} \ .
\label{eq:2.12}
\end{equation}
Then the constraints take the more pleasing form (cf.~\cite{Ash1,Kod,Barb})
\begin{equation}
\label{eq:2.13}
\tilde{{\cal H}}_0^{{\rm ADM}} \equiv
\tilde{{\cal H}}_0 \mp i \, \partial_i \bigl ( e^i \,\!_a \, \tilde{{\cal J}}_a \bigr ) \stackrel{!}{=} 0 \ ,
\end{equation}
with
\begin{equation}
\label{eq:2.13a}
\tilde{{\cal H}}_0={\frac{1}{\gamma}} \, e_{i a}
\Bigl [ \tilde{\varepsilon}^{i j k} \, {\cal F}_{j k a}+
{\textstyle \frac{2}{3}} \, \Lambda \, {\tilde e}^i \,\!_{a} \Bigr ] \ ,
\end{equation}
\begin{equation}
\label{eq:2.14}
\tilde{{\cal H}}_i \equiv
\mp \frac{2 i}{\gamma} \,
\Bigl [ {\tilde{e}}^j \,\!_a \, \partial_j {\cal A}_{i a}-
 {\tilde{e}}^j \,\!_a \, \partial_i {\cal A}_{j a}+ {\cal A}_{i a} \, \partial_j {\tilde{e}}^j \,\!_a \Bigr ]
 \stackrel{!}{=} 0 \ ,
\end{equation}
\begin{equation}
\tilde{{\cal J}}_a \equiv
\pm \frac{2 i}{\gamma} \,
\Bigl [ \partial_i {\tilde{e}}^i \,\!_a+
 \varepsilon_{a b c} \, {\tilde{e}}^i\,\!_c \, {\cal A}_{i b} \Bigr ]  \stackrel{!}{=} 0 \ ,
\label{eq:2.15}
\end{equation}
and the Hamiltonian
\begin{equation}
H= \int d^3 x \left (N \tilde{{\cal H}}_0+N^i \tilde{{\cal H}}_i+\Omega_a \tilde{{\cal J}}_a \right )\ ,
\label{eq:2.15a}
\end{equation}
endowed with the symplectic structure (\ref{eq:2.18a}), (\ref{eq:2.18b}),
is dynamically equivalent to the ADM-Hamiltonian (\ref{eq:2.8}).
In fact, as long as $\Lambda \not= 0$, the constraints (\ref{eq:2.13a}) -
(\ref{eq:2.15}) can all be expressed in terms
of the single tensor density $\tilde{{\cal G}}^i_{\Lambda , a}$ defined by
\begin{equation}
\tilde{{\cal G}}^i_{\Lambda , a}= {\textstyle \frac{1}{2}} \, \tilde{\varepsilon}^{i j k} \, 
{\cal F}_{j k a} + {\textstyle \frac{1}{3}} \, \Lambda \, \tilde{e}^i \,\!_a \ ,
\label{eq:2.16}
\end{equation}
namely
\begin{equation}
\tilde{{\cal H}}_0 \equiv
\frac{2}{\gamma} \, e_{i a} \, \tilde{{\cal G}}^i_{\Lambda , a} \stackrel{!}{=} 0 \ ,
\label{eq:2.17-}
\end{equation}
\begin{equation}
\label{eq:2.17}
\tilde{{\cal H}}_i =
\pm \frac{2 i}{\gamma} \,
\, \mbox{\raisebox{-1.9 ex}{$\tilde{}$} \hspace{-1.2 ex}} \varepsilon_{i j k} 
\, \tilde{e}^j \,\!_a \, \tilde{{\cal G}}^k_{\Lambda , a} -
{\cal A}_{i a} \, \tilde{{\cal J}}_a \stackrel{!}{=} 0 \ ,
\end{equation}
\begin{equation}
\tilde{{\cal J}}_a  = 
\pm \frac{6 i}{\gamma \Lambda} \, {\cal D}_i \, \tilde{{\cal G}}^i_{\Lambda , a}
\stackrel{!}{=} 0 \ ,
\label{eq:2.17+}
\end{equation}
where ${\cal D}_i$ is the covariant derivative with respect to
the connection ${\cal A}_{i a}$. For $\Lambda=0$ the relation of the
constraint $\tilde{{\cal J}}_a$ with $\tilde{{\cal G}}^i_{\Lambda , a}$ is lost. 
A simple way to satisfy
all the constraints (\ref{eq:2.17-})-(\ref{eq:2.17+}) for $\Lambda \not= 0$
is to restrict the phase space
by the nine conditions
\begin{equation}
\tilde{{\cal G}}^i_{\Lambda , a} \stackrel{!}{=} 0 \ .
\label{eq:2.18}
\end{equation}
Eqs.~(\ref{eq:2.18}) are more restrictive than the seven
eqs.~(\ref{eq:2.17-})-(\ref{eq:2.17+}) which they imply, i.e. we can only hope to get special
solutions in this manner. Remarkably, eqs.~(\ref{eq:2.18}),
if imposed as initial conditions, remain satisfied for all times
under the time evolution generated by the Hamiltonian (\ref{eq:2.15a}).
This follows from the Poisson-brackets
\begin{equation}
\left \{
\int d^3 x \, N^i \tilde{{\cal H}}_i \, , \, \int d^3 y \, \lambda_{j a} \, \tilde{{\cal G}}^j_{\Lambda , a}
\right \}
=
\int d^3 z \,
\left ( N^i \partial_i \lambda_{j a}+\lambda_{i a} \partial_j N^i \right )
\tilde{{\cal G}}^j_{\Lambda , a} \ ,
\end{equation}
\begin{equation}
\left \{
\int d^3 x \, \Omega_a \, \tilde{{\cal J}}_a \, , \, \int d^3 y \, \lambda_{j b} \, 
\tilde{{\cal G}}^j_{\Lambda , b}
\right \}
=
\int d^3 z \, \Omega_a \, \varepsilon_{a b c} \, \lambda_{j b} \,
\tilde{{\cal G}}^j_{\Lambda , c} \ ,
\end{equation}
\begin{equation}
\label{eq:2.19}
\left \{
\int d^3 x \, N  \tilde{{\cal H}}_0 \, , \, \int d^3 y \, \lambda_{j a} \, 
\tilde{{\cal G}}^j_{\Lambda , a} \right \}
=
\pm {\textstyle \frac{i}{2}} \int d^3 z \, \frac{N}{\sqrt{h}}
 \left ( e_{i a} e_{j b} - 2 \, e_{i b} e_{j a} \right ) \tilde{\varepsilon}^{j k \ell} \, {\cal D}_k
 \lambda_{\ell b} \, \tilde{{\cal G}}^i_{\Lambda , a} \ ,
\end{equation}
which may be verified with some labor using eqs.~(\ref{eq:2.16}) and
(\ref{eq:2.17-})-(\ref{eq:2.17+}). They
imply that on the subspace $\tilde{{\cal G}}^i_{\Lambda , a}=0$ of phase-space
\begin{equation}
\left \{ H \, , \, \tilde{{\cal G}}^i_{\Lambda , a} \right \} = 0 \ ,
\label{eq:2.20}
\end{equation}
i.e. this subspace is conserved. 

The equations (\ref{eq:2.18}) bear
a superficial formal similarity to Einstein's field equations
\begin{equation}
 G^\mu_{\Lambda , \nu} :={G^\mu}_{\nu} + \Lambda \, \delta^\mu_{\nu}=0
\label{eq:2.21}
\end{equation}
in 4 space-time dimensions ($\mu,\nu=0,1,2,3$) with the 4-dimensional
Einstein-tensor
${G^\mu}_{\nu}$ satisfying the Bianchi-identity
\begin{equation}
\nabla_{\! \mu} \, {G^\mu}_{\nu} \equiv 0
\label{eq:2.21a}
\end{equation}
and also $\nabla_{\! \mu} \, G^\mu_{\Lambda ,\nu}=0$, because the affine connection satisfies
the metric postulate. Since $\tilde{{\cal G}}^i_{\Lambda , a}$ similarly decomposes
in a curvature part satisfying a Bianchi-identity
\begin{equation}
{\cal D}_i \left ( \tilde{\varepsilon}^{i j k} \, {\cal F}_{j k a} \right ) \equiv 0 \ ,
\label{eq:2.21b}
\end{equation}
and a cosmological term proportional to $\Lambda$ it is a three-dimensional
analog of $G^\mu_{\Lambda , \nu}$. The analogy extends even to
${\cal D}_i \, \tilde{{\cal G}}^i_{\Lambda , a}=0$, which holds due to the Bianchi-identity
but requires in addition for the constraint (\ref{eq:2.15}). However, it has to
be kept in mind that the
spin-connection ${\cal A}_{i a}$ and the densitized inverse triad $\tilde{e}^i \,\!_{a}$
in $\tilde{{\cal G}}^i_{\Lambda , a}$ are still {\it independent} variables. The equations
(\ref{eq:2.18}) therefore are not a closed set of field equations on the
spatial manifolds.


\section{Quantization}
\label{section3}

Canonical quantization in the triad representation is achieved by imposing
the commutation relations
\begin{equation}
\left [ \tilde{e}^i \,\!_a (x) \, , \, p_{j b} (y) \right ]
=
i \hbar \, \delta^i_j \, \delta_{a b} \, \delta^3 (x-y)
\label{eq:3.1}
\end{equation}
and representing $p_{i a}(x)$ as
\begin{equation}
p_{i a}
=
\frac{\hbar}{i} \, \frac{\delta}{\delta \tilde{e}^i \,\!_{a}(x)} \ .
\label{eq:3.2}
\end{equation}
This implies for the ${\cal A}_{i a}$ the representation
\begin{equation}
{\cal A}_{i a}(x)
=
\omega_{i a} (x) \pm \frac{\gamma \hbar}{2} \,
\frac{\delta}{\delta \tilde{e}^i \,\!_{a}(x)} \ ,
\label{eq:3.3}
\end{equation}
where $\omega_{i a} (x)$, given by (\ref{eq:2.19a}), (\ref{eq:2.20a}) is a
functional of $\tilde{e}^j \,\!_{b} (y)$ and a diagonal operator in this
representation.
We now have to choose a special factor ordering  in the constraint
operators $\tilde{{\cal J}}_a$, $\tilde{{\cal H}}_i$ and $\tilde{{\cal H}}_0$. 
It turns out that $\tilde{{\cal J}}_a$
does not suffer from an ordering ambiguity. We choose the factor ordering in
$\tilde{{\cal H}}_0$
and  $\tilde{{\cal H}}_i$ as given in (\ref{eq:2.13a}) and (\ref{eq:2.14}) in order to achieve
closure of the algebra of the generators. Explicitly, the generators are
then given by eqs.~(\ref{eq:2.13a})-(\ref{eq:2.15}) or eqs.~(\ref{eq:2.17-})-(\ref{eq:2.17+})
with the
ordering of $\tilde{e}^i \,\!_{a}$, ${\cal A}_{j b}$ given there. The algebra of the
infinitesimal generators is obtained as\footnote{The algebra of the constraint operators
has been discussed intensively in the literature, see e.g.~\cite{Ash1,Kod,JacSmo}. The
factor ordering and the corresponding operator algebra considered here are in agreement
with Ashtekar's results in \cite{Ash1}.} 
\begin{equation}
\label{eq:3.0.1}
\left [
\int d^3 x \, \xi^i \tilde{{\cal H}}_i \, , \, \int d^3 y \, \varphi_a \, \tilde{{\cal J}}_a
\right ]
=
i \hbar \int d^3 z\,\left ( \xi^i \, \partial_i \varphi_a \right ) \tilde{{\cal J}}_a \ ,
\end{equation}
\begin{equation}
\label{eq:3.0.2}
\left [
\int d^3 x \, \xi^i \tilde{{\cal H}}_i \, , \, \int d^3 y \, \eta^j \tilde{{\cal H}}_j
\right ]
=
i \hbar \int d^3 z \, \left ( \xi^i \, \partial_i \eta^j - \eta^i \, \partial_i \xi^j \right )
\tilde{{\cal H}}_j \ ,
\end{equation}
\begin{equation}
\left [
\int d^3 x \, \varphi_a \, \tilde{{\cal J}}_a \, , \, \int d^3 y \, \psi_b \, \tilde{{\cal J}}_b
\right ]
=
i \hbar \int d^3 z \, \varepsilon_{a b c} \, \varphi_a \, \psi_b \, \tilde{{\cal J}}_c \ ,
\label{eq:3.0.3}
\end{equation}
\begin{equation}
\label{eq:3.0.4}
\left [
\int d^3 x \, \varphi_a \, \tilde{{\cal J}}_a \, , \, \int d^3 y \, N \tilde{{\cal H}}_0
\right ]
= 0 \ ,
\end{equation}
\begin{equation}
\label{eq:3.0.5}
\left [
\int d^3 x \, \xi^i \tilde{{\cal H}}_i \, , \, \int d^3 y \, N \tilde{{\cal H}}_0
\right ]
=
i \hbar \int \, d^3 z \, \left ( \xi^i \, \partial_i N \right ) \tilde{{\cal H}}_0 \ ,
\end{equation}
\begin{equation}
\label{eq:3.0.6}
\left [
\int d^3 x \, N \tilde{{\cal H}}_0 \, , \, \int d^3 y \, M \tilde{{\cal H}}_0
\right]
=
i \hbar \int d^3 z \, \left ( N \partial_i M -M \partial_i N \right ) h^{i j}
\left ( \tilde{{\cal H}}_j + {\cal A}_{j a} \, \tilde{{\cal J}}_a \right) \ .
\end{equation}
On the right hand side of these equations all generators appear on
the right, which means that the algebra closes, at least formally (i.e. in the absence of
any regularization procedure), without any quantum
corrections.

Following Dirac \cite{Dirac}, physical states $\Psi [ \tilde{e}^i \,\!_a ]$
must satisfy
\begin{eqnarray}
\label{eq:3.4}
\tilde{{\cal J}}_a \, \Psi [ \tilde{e}^i \,\!_a ]  & \stackrel{!}{=} 0 & \qquad 
\mbox{Lorentz invariance} \ ,  \\[2 ex]
\label{eq:3.5}
\tilde{{\cal H}}_i \, \Psi [ \tilde{e}^i \,\!_a ] & \stackrel{!}{=} 0 & \qquad
\mbox{diffeomorphism invariance} \ , \\[2 ex] 
\label{eq:3.6} 
\tilde{{\cal H}}_0 \, \Psi [ \tilde{e}^i \,\!_a ] & \stackrel{!}{=} 0 & \qquad
\mbox{time-redefinition invariance} \ .
\end{eqnarray}
Moreover, since the Lorentz constraint (\ref{eq:3.4}) guarantees only invariance under
local $SO(3)$-gauge-transformations of the triad $\tilde{e}^i \,\!_a$, 
while the full symmetry group is given by
$O(3)$, we further have to impose a discrete, global parity requirement
\begin{equation}
  \label{eq:3.6+}
  {\cal P} \, \Psi [ \tilde{e}^i \,\!_a ] := \Psi [- \tilde{e}^i \,\!_a ] \stackrel{!}{=} 
  + \Psi [ \tilde{e}^i \,\!_a ]  \ ,
\end{equation}
where ${\cal P}$ denotes the parity operator acting on functionals of the triad.

As in the classical theory, the constraints (\ref{eq:3.4})-(\ref{eq:3.6}) 
on physical states  are all
satisfied if the stronger conditions
\begin{equation}
\tilde{{\cal G}}^i_{\Lambda , a} \, \Psi [ \tilde{e}^i \,\!_a ] \stackrel{!}{=} 0
\label{eq:3.7}
\end{equation}
hold, where $\tilde{{\cal G}}^i_{\Lambda , a}$ is the tensor density defined by
eqs.~(\ref{eq:2.16}), (\ref{eq:2.12}) in terms of the operators
$\tilde{e}^i \,\!_{a}$ and ${\cal A}_{i a}$ given by eqs.~(\ref{eq:2.4}),
(\ref{eq:2.11}). Remarkably, the quantum operators $\tilde{{\cal G}}^i_{\Lambda , a}$
turn out to commute among themselves.
It can be
seen from eqs.~(\ref{eq:2.17-})-(\ref{eq:2.17+}), which must now be read as operator
equations, that eqs.~(\ref{eq:3.4})-(\ref{eq:3.6}) are implied by
(\ref{eq:3.7}). The subspace of physical states satisfying
(\ref{eq:3.7}) is the quantum version of the invariant subspace of
classical phase-space defined by eqs.~(\ref{eq:2.18}).

To find the
solutions of eqs.~(\ref{eq:3.7}) it is useful to proceed in two steps.
First, it is convenient to perform a similarity transformation (cf.~\cite{Kod})
\begin{equation}
\Psi = \exp \left [ \mp \frac{2}{\gamma \hbar} \, \phi \right ] \cdot \Psi' \ ,
\label{eq:3.8}
\end{equation}
where $\phi$ was defined in eq.~(\ref{eq:2.20a}). Under this
transformation, the operators ${\cal A}_{i a}$ according to (\ref{eq:3.3})
transform like
\begin{equation}
\exp \left [ \pm \frac{2}{\gamma \hbar} \, \phi \right ] \cdot 
{\cal A}_{i a} \cdot \exp \left [ \mp \frac{2}{\gamma \hbar} \, \phi \right ]
= \pm \frac{\gamma \hbar}{2} \, \frac{\delta}{\delta \tilde{e}^i \,\!_{a}} \ ,
\label{eq:3.9}
\end{equation}
and eq.~(\ref{eq:3.7}) becomes explicitly
\begin{equation}
\left [
\tilde{\varepsilon}^{i m n}
\left (
\pm \gamma \hbar \, \partial_m \frac{\delta}{\delta \tilde{e}^n \,\!_{a}} +
 \frac{\gamma^2 \hbar^2}{4} \, \varepsilon_{a b c} \,
\frac{\delta^2}{\delta \tilde{e}^m \,\!_{b} \, \delta \tilde{e}^n \,\!_{c}}
\right ) + \frac{2 \Lambda}{3} \, \tilde{e}^i \,\!_{a}
\right ] \, \Psi' = 0 \ .
\label{eq:3.10}
\end{equation}
\\
As a second step, we now consider a representation of
$\Psi' [ \tilde{e}^i \,\!_{a} ]$ by a generalized Fourier integral
\begin{equation}
\Psi' [ \tilde{e}^i \,\!_{a} ]
=
\int_{\Gamma} {\cal D}^9 [ {\cal A}_{i a} ] \,
\exp
\left [ \pm \frac{2}{\gamma \hbar} \int d^3 x \, \tilde{e}^i \,\!_a \, {\cal A}_{i a} \right ]
\cdot \tilde{\Psi} [{\cal A}_{i a} ]
\label{eq:3.11}
\end{equation}
\\
where the complex integration manifold $\Gamma$ is chosen
in such a way that partial integrations with respect to ${\cal A}_{i a}$
are permitted without any boundary terms. Besides these restrictions,
$\Gamma$ may be chosen arbitrarily to guarantee the existence of the
functional integral (\ref{eq:3.11}) (cf.~discussions of the
homogeneous Bianchi~IX model \cite{Pat1,Pat3}). 
Different choices of $\Gamma$ within these
restrictions, which cannot be deformed into each other continuosly
without crossing a singularity of the integrand, will, in general,
correspond to different solutions. Under the transformation
(\ref{eq:3.11}) the fundamental operators ${\cal A}_{i a}$, $\tilde{e}^i \,\!_{a}$
transform like
\begin{equation}
\tilde{e}^i \,\!_a \cdot \Psi'
\ \mapsto \ 
\mp \frac{\gamma \hbar}{2} \, \frac{\delta \tilde{\Psi}}{\delta {\cal A}_{i a}} \qquad , \qquad
\label{eq:3.12}
\frac{\delta \Psi'}{\delta \tilde{e}^i \,\!_{a}} \ \mapsto 
\ \pm \frac{2}{\gamma \hbar} \, {\cal A}_{i a} \cdot \tilde{\Psi} \quad ,
\end{equation}
and equation (\ref{eq:3.10}) becomes
\begin{equation}
\left [
\tilde{\varepsilon}^{i j k} \, {\cal F}_{j k a} \mp \frac{\gamma \hbar \Lambda}{3} \,
\frac{\delta}{\delta {\cal A}_{i a}}
\right ] \tilde{\Psi} = 0 \ .
\label{eq:3.14}
\end{equation}
\\
Up to a normalization factor ${\cal N}$, the unique solution of (\ref{eq:3.14}) is the 
Chern-Simons state (cf.~\cite{Kod}) 
\begin{equation}
\tilde{\Psi}_{{\rm CS}} [ {\cal A}_{i a} ] = {\cal N} \,
\exp \left [ \pm \frac{3}{\gamma \hbar \Lambda} \, {\cal S}_{{\rm CS}} [ {\cal A}_{i a} ] \right ]
\label{eq:3.15}
\end{equation}
\\
with the Chern-Simons functional
\begin{equation}
{\cal S}_{{\rm CS}} [ {\cal A}_{i a} ]
=
\int d^3 x \, \tilde{\varepsilon}^{i j k}
\left ( {\cal A}_{i a} \, \partial_j {\cal A}_{k a}+{\textstyle \frac{1}{3}} \, \varepsilon_{a b c} \,
{\cal A}_{i a} \, {\cal A}_{j b} \, {\cal A}_{k c} \right ) \ .
\label{eq:3.16}
\end{equation}
\\
In the $\tilde{e}^i \,\!_{a}$-representation the corresponding wavefunctional is given by
\begin{equation}
\Psi_{{\rm CS}} [ \tilde{e}^i \,\!_a ]
= {\cal N}
\int_{\Gamma} {\cal D}^9 [ {\cal A}_{i a} ] \, \exp
\left [
\pm \frac{1}{\gamma \hbar}
\left (
\int d^3 x \, \tilde{\varepsilon}^{i jk} \, e_{i a} {\cal D}_j e_{k a} + \frac{3}{\Lambda} \, 
{\cal S}_{{\rm CS}} [ {\cal A}_{i a} ]
\right )
\right ] \ .
\label{eq:3.17}
\end{equation}
\\
The state (\ref{eq:3.17}) is obviously diffeomorphism-invariant, and it is also
gauge-invariant under sufficiently small
gauge-transformations (i.e. those which are continuously connected to the identical
transformation):\footnote{Here and in the following, we shall refer to the 
$SO(3)$-gauge-invariance just as ``gauge-invariance'' for short. The 
diffeo\-mor\-phism- and the time-redefinition-invariance, which are of course 
inherent gauge-symmetries of the theory as well, will allways be mentioned separately.} 
The contribution from the similarity transformation (\ref{eq:3.9}) and the
Fourier term from (\ref{eq:3.11}) fit perfectly together to give the
first gauge-invariant term in the exponent of (\ref{eq:3.17}), while the
second term proportional to ${\cal S}_{{\rm CS}}$ is a well-known
gauge-invariant functional. The wavefunctional $\Psi_{{\rm CS}} [ \tilde{e}^i \,\!_a ]$ given in
(\ref{eq:3.17}) further turns out to be parity invariant, as it was required by the
condition (\ref{eq:3.6+}).

However, for a trivial choice of the prefactor ${\cal N}$ in (\ref{eq:3.17}) the state
$\Psi_{{\rm CS}} [ \tilde{e}^i \,\!_a ]$ \emph{fails} to be invariant under \emph{large}
gauge-transformations of the triad, since the Chern-Simons functional in (\ref{eq:3.17}) 
transforms non-trivially under such
transformations (cf.~\cite{Ash1,Wein2}). At this point it is helpful to notice that
the prefactor ${\cal N}$ in (\ref{eq:3.17}), underlying the only restriction
\begin{equation}
  \label{eq:3.17+}
  \frac{\delta {\cal N}}{\delta \tilde{e}^i \,\!_a} \stackrel{!}{=} 0 \ ,
\end{equation}
is just required to be constant under \emph{infinitesimal} variations of $\tilde{e}^i \,\!_a$, 
while it may still depend on topological invariants of the triad. In 
section~\ref{section4.2.0}
we will make use of this remarkable freedom, choosing the normalization factor ${\cal N}$
in such a way that the state $\Psi_{{\rm CS}} [ \tilde{e}^i \,\!_a ]$ becomes invariant even under
large gauge-transformations of the triad with a non-trivial winding number.

Unfortunately,
the integration manifold $\Gamma$ in eq.~(\ref{eq:3.17}) can not be given explicitly,
but we will argue that several topologically inequivalent choices for 
$\Gamma$ do exist, which give rise to
linearly independent quantum states $\Psi_{{\rm CS}} [ \tilde{e}^i \,\!_{a} ]$. These
different states in the $\tilde{e}^i \,\!_a$-representation  
arise all from the \emph{one} Chern-Simons state in the
${\cal A}_{i a}$-representation, a phenomenon which is well-known from discussions of 
the homogeneous Bianchi~IX model in earlier papers \cite{Pat1,Pat3}. Together
these states span the subspace of physical states corresponding to the invariant
subspace of phase-space defined classically by $\tilde{{\cal G}}^i_{\Lambda , a} =0$.


\section{Asymptotic expansions of the Chern-Simons state}
\label{section4}

Since the functional integral occuring in the $\tilde{e}^i \,\!_a$-representation of the
Chern-Simons state (\ref{eq:3.17}) is too complicated to be performed analytically,
we will restrict ourselves to an asymptotic evaluation of the wavefunctional (\ref{eq:3.17})
in several interesting parameter regimes. The possible different asymptotic regimes can be
displayed by rewriting the Chern-Simons state (\ref{eq:3.17}) in 
dimensionless quantities. Therefore, we introduce the three
fundamental length-scales of the theory, namely the Planck-scale
\begin{equation}
  \label{eq:4.1.1}
  a_{\rm Pl} := \sqrt{\gamma \hbar} \ ,
\end{equation}
the cosmological scale parameter
\begin{equation}
  \label{eq:4.1.2}
  a_{\rm cos} := \textstyle{\left ( \int d^3 x \, \sqrt{h} \right  )^{1/3}}
\end{equation}
and a third  length-scale, which is associated with the cosmological constant $\Lambda$:
\begin{equation}
  \label{eq:4.1.3}
  a_{\Lambda} := \sqrt{\frac{3}{\Lambda}} \ .
\end{equation}
These three length-scales give rise to the definition of two dimensionless parameters,
for example
\begin{equation}
  \label{eq:4.1.4}
  \kappa := \left ( \frac{a_{\rm cos}}{a_{\Lambda}} \right )^2 = \frac{\Lambda}{3} \, a_{\rm cos}^2
\qquad , \qquad \mu := \left ( \frac{a_{\rm cos}}{a_{\rm Pl}} \right )^2 =
\frac{a_{\rm cos}^2}{\gamma \hbar} \  \ .
\end{equation}
Moreover, we may rescale the triad fields with the help of the cosmological scale 
parameter $a_{\rm cos}$ to arrive at dimensionless field 
variables denoted by a prime:\footnote{
By definition, the Ashtekar variables ${\cal A}_{i a}$ carry no dimension and need not
to be rescaled.}
\begin{equation}
  \label{eq:4.1.5}
  e'_{i a} = a_{\rm cos}^{-1} \, e_{i a} \qquad , \qquad {\tilde{e}}'^i \,\!_{a} =a_{\rm cos}^{-2} \, 
 {\tilde{e}^i} \,\!_{a} \qquad , \qquad
 \sqrt{h'}=a_{\rm cos}^{-3} \, \sqrt{h} \ \ .
\end{equation}
Making use of eqs.~(\ref{eq:4.1.1})-(\ref{eq:4.1.5})
the Chern-Simons state (\ref{eq:3.17}) reduces to the form
\begin{equation}
  \label{eq:4.2.1}
  \Psi_{\rm CS} [ \tilde{e}^i \,\!_a ] = {\cal N} \int_{\Gamma} {\cal D}^9 [ {\cal A}_{i a} ] \,
   \exp \Big [ \pm \mu F  \, \Big ] \ ,
\end{equation}
where the exponent $F$ is defined by
\begin{equation}
  \label{eq:4.2.2}
  F := \int d^3 x \, \tilde{\varepsilon}^{i j k} \, e'_{i a} \, {\cal D}_j e'_{k a} +
  \frac{1}{\kappa} \, {\cal S}_{\rm CS} [ {\cal A}_{i a} ] \ .
\end{equation}


\subsection{The semiclassical limit $\mu \to \infty$}
\label{section4.1}

Because of the Gaussian saddle-point form of (\ref{eq:4.2.1}) with respect to the
parameter $\mu$ it is natural to study the limit $\mu \to \infty$ first. This limit
corresponds to the regime $a_{\rm cos} \gg a_{\rm Pl}$, and also to the formal limit
$\hbar \to 0$ (cf.~eq.~(\ref{eq:4.1.4})), so we shall refer to it as the \emph{semiclassical} 
limit for short.
In the limit $\mu \to \infty$ the asymptotic form of the integral (\ref{eq:4.2.1}) becomes in
leading order of $\mu$
\begin{equation}
\Psi_{\rm CS } [ \tilde{e}^i \,\!_{a} ] \stackrel{\mu \to \infty}{\propto}
{\cal N} \, \left |
\frac{\mu \, \delta^2 {F}}{\delta {\cal A}_{i a} (x) \, \delta {\cal A}_{j b} (y)}
\right |^{-\frac{1}{2}} \cdot \exp \Big [ \pm \mu F \, \Big ] \ ,
\label{eq:4.2.3}
\end{equation}
\\
where an infinite prefactor in (\ref{eq:4.2.3}) has been omitted.
The asymptotic expression (\ref{eq:4.2.3}) has to be evaluated 
at a saddle-point of the exponent $F$ with
respect to ${\cal A}_{i a}$, which is obtained by solving the saddle-point equations
\begin{equation}
\frac{\delta F}{\delta {\cal A}_{i a}} = 2 \, \tilde{e}'^i \,\!_{a} +
 \frac{1}{\kappa} \, \tilde{\varepsilon}^{i j k} \, {\cal F}_{j k a} \stackrel{!}{=} 0 \ .
\label{eq:4.3}
\end{equation}
The equations (\ref{eq:4.3}) more explicitly take the form
\begin{equation}
\tilde{\varepsilon}^{i j k} \left ( \partial_j {\cal A}_{k a} +
 {\textstyle \frac{1}{2}} \, \varepsilon_{a b c} \, {\cal A}_{j b} \, {\cal A}_{k c} \right )
=
- \frac{\Lambda}{3} \, \tilde{e}^i \,\!_{a} \ ,
\label{eq:4.4}
\end{equation}
and coincide with the classical equations $\tilde{{\cal G}}^i_{\Lambda , a}=0$
as they should, since the latter constitute the classical limit of the gravitational
Chern-Simons state.
The saddle-point equations (\ref{eq:4.4}) must be read as 
determining implicitly the complex spin-connection
${\cal A}_{i a}$ for any given real triad $\tilde{e}^i \,\!_{a}$, for which we wish to
evaluate $\Psi_{\rm CS} [ \tilde{e}^i \,\!_{a} ]$. Since $\tilde{e}^i \,\!_{a}$
carries information
about the coordinate system and the local $SO(3)$-gauge-degrees of freedom,
the solutions ${\cal A}_{i a}$ of (\ref{eq:4.4}) for a \emph{given} triad
$\tilde{e}^i \,\!_{a}$ have \emph{no} further gauge-freedom. This is why
we expect a discrete, finite set of solutions ${\cal A}_{i a}$ of (\ref{eq:4.4})
for a fixed triad $\tilde{e}^i \,\!_{a}$. A detailed mathematical discussion of 
the solvability properties of the semiclassical saddle-point equation (\ref{eq:4.4}) will
be given in appendix~\ref{appA.1}.

For a fixed triad $\tilde{e}^i \,\!_a$ the number of the different gauge-fields
${\cal A}_{i a}$ solving (\ref{eq:4.4}) will
depend on the topology of the spatial manifold ${\cal M}_3$: For example,
if ${\cal M}_3$ has the topology of the 3-sphere $S^3$, five distinct solutions ${\cal A}_{i a}$
of the corresponding saddle-point equations are found for spatially
\emph{homogeneous} 3-manifolds, which are described by the Bianchi~IX model
(cf.~\cite{Pat1,Pat3}). It follows from the arguments given in appendix~\ref{appA.1},
that this number of saddle-points is
preserved under sufficiently small inhomogeneous perturbations
of the triad  $\tilde{e}^i \,\!_{a}$. We therefore find \emph{five}
physically inequivalent solutions ${\cal A}_{i a}$ in this case. If we consider
manifolds ${\cal M}_3$ with the topology of the 3-torus $T^3$, the subset of homogeneous
manifolds is described by the Bianchi~I model, restricting the number of
independent solutions ${\cal A}_{i a}$ of (\ref{eq:4.4}) to be \emph{two}, as in this
homogeneous model. Considering other topologies of ${\cal M}_3$, the number of
inequivalent saddle-points will differ further. However, we will see in
subsection~\ref{section4.2} that, for \emph{any} given topology of the spatial
3-manifold ${\cal M}_3$, the
number of distinct saddle-points ${\cal A}_{i a}$ of (\ref{eq:4.4}) should
\emph{at least} be two. 

Given a topology of ${\cal M}_3$ and a saddle-point solution
${\cal A}_{i a}$ of (\ref{eq:4.4}), the evaluation of (\ref{eq:4.2.3}) at this
saddle-point gives a possible semiclassical contribution to the
Chern-Simons state $\Psi_{\rm CS} [ \tilde{e}^i \,\!_a ]$ 
in  the limit $\mu \to \infty$. It will depend on the choice of the 
integration contour $\Gamma$ in (\ref{eq:4.2.1}) whether this particular 
saddle-point contributes to the functional integral or not.
Under gauge- or coordinate-
transformations of the triad $\tilde{e}^i \,\!_{a}$ the fixed solution
${\cal A}_{i a}$ of (\ref{eq:4.4}) transforms like a spin-connection, since
(\ref{eq:4.4}) is a coordinate- and gauge-covariant equation. Consequently, the
semiclassical expression (\ref{eq:4.2.3}) remains unchanged under (sufficiently small)
gauge-transformations, as it indeed must be the case, since $\Psi_{\rm CS}$, also for
$\mu \to \infty$, was constructed as as a coordinate- and gauge-invariant state.
Therefore, we may solve the equations (\ref{eq:4.4}) in any desired gauge for
 $\tilde{e}^i \,\!_{a}$, fixing automatically a gauge for the solutions ${\cal A}_{i a}$.

Any \emph{possible} saddle-point contribution (\ref{eq:4.2.3}) for a given saddle-point
${\cal A}_{i a}$ can be chosen to become the \emph{dominant} contribution to the functional
integral in (\ref{eq:4.2.1}) in the limit $\mu \to \infty$ by choosing the complex
integration manifold $\Gamma$ suitably.
So the number of linearly independent semiclassical wavefunctionals
$\Psi_{\rm CS} [ \tilde{e}^i \,\!_{a} ]$ equals the number of inequivalent
saddle-points ${\cal A}_{i a}$ of (\ref{eq:4.4}). This is also the number of linearly
independent \emph{exact} wavefunctionals $\Psi_{\rm CS} [ \tilde{e}^i \,\!_{a} ]$, because
the complex integration manifold $\Gamma$, constructed as a contour of steepest
descend to a given saddle-point ${\cal A}_{i a}$, satisfies the requirements for $\Gamma$
in eq.~(\ref{eq:3.11}) and may therefore be used to define an exact
wavefunctional (\ref{eq:4.2.1}). We conclude that the \emph{one} Chern-Simons
state (\ref{eq:3.15}) in the complex Ashtekar representation generates a
discrete, finite set of linearly independent gravitational states in the real
triad representation, which differ by the topology of the integration manifolds
$\Gamma$ connecting the two representations via (\ref{eq:3.11}). The number of
the different Chern-Simons states in the $\tilde{e}^i \,\!_{a}$-representation depends on the
topology of the spatial manifold ${\cal M}_3$ and should at least be \emph{two}.

We will now try to construct explicit solutions ${\cal A}_{i a}$ 
of the non-linear, partial differential equations
(\ref{eq:4.4}). In general, analytical solutions of this complicated set of
equations are not available, so we will 
restrict ourselves to asymptotic solutions in the two different
limits $\kappa \to \infty$ and $\kappa \to 0$, which will be treated in sections~\ref{section4.2}
and \ref{section4.3}, respectively.


\subsection{The limit of large scale parameter $\mu \to \infty , \kappa \to \infty$}
\label{section4.2}

According to our definition of the parameters $\mu$ and $\kappa$ in eq.~(\ref{eq:4.1.4}),
the limit $\kappa \to \infty$ within the semiclassical limit $\mu \to \infty$
can be realized by taking the scale parameter $a_{\rm cos}$ of the spatial manifold
sufficiently large,
$a_{\rm cos} \gg a_{\rm Pl} , a_{\Lambda}$.
In this special asymptotic regime, solutions of (\ref{eq:4.4})
can be found by inserting the ansatz
\begin{equation}
{\cal A}_{i a} \stackrel{\kappa \to \infty}{\sim} \sqrt{\kappa} \,  c_{i a}^{(0)} + 
{\cal O} ( \kappa^0 ) 
\label{eq:4.9}
\end{equation}
into the saddle-point equations
\begin{equation}
  \label{eq:4.9+}
  \tilde{\varepsilon}^{i j k} \left ( \partial_j {\cal A}_{k a} + {\textstyle \frac{1}{2}} \,
  \varepsilon_{a b c} \, {\cal A}_{j b} \, {\cal A}_{k c} \right )= - \kappa \, \tilde{e}'^{i} \,\!_a \ .
\end{equation}
Then we find the two solutions
\begin{equation}
c_{i a}^{(0)}=\pm i  \, e'_{i a} \ ,
\label{eq:4.10}
\end{equation}
or, equivalently,
\begin{equation}
  \label{eq:4.10+}
  {\cal A}_{i a} \stackrel{\kappa \to \infty}{\sim} \pm i \, \sqrt{\frac{\Lambda}{3}} \, e_{i a} +
  {\cal O} ( \kappa^0) \ . 
\end{equation}
We
should stress that the two signs occurring in (\ref{eq:4.10}), (\ref{eq:4.10+})
are \emph{independent} of the double sign in (\ref{eq:2.11}), i.e. for
both possible definitions (\ref{eq:2.11}) of the Ashtekar variables we
find two independent, complex conjugate solutions ${\cal A}_{i a}$ of the
saddle-point equations (\ref{eq:4.9+}) in the limit $\kappa \to \infty$, 
corresponding to two semiclassical
wavefunctions via (\ref{eq:4.2.3}). To avoid confusion, we will 
discuss only one of these solutions in the following, which is obtained by choosing the
upper sign in eqs.~(\ref{eq:4.10}), (\ref{eq:4.10+}). The corresponding results for
the second solution may then be obtained at any time by a complex conjugation.

The result (\ref{eq:4.10+}) can be improved by calculating the coefficients
$c_{i a}^{(n)}$ of the asymptotic series 
\begin{equation}
{\cal A}_{i a} \stackrel{\kappa \to \infty}{\sim} 
\sum_{n=0}^\infty c_{i a}^{(n)} \, \kappa^{(1-n)/2} \ .
\label{eq:4.11}
\end{equation}
All coefficients in (\ref{eq:4.11})
can be calculated analytically, since, in any order of $\kappa$, the non-Abelian
term in (\ref{eq:4.9+}) contains the unknown coefficient $c_{i a}^{(n)}$, while
the non-local term in (\ref{eq:4.9+}) is known from the previous orders.
Consequently, the recursion equations determining $c_{i a}^{(n)}$ are just
algebraic equations at each space-point, which, moreover, are linear and
analytically solvable for $n>0$. The first three terms of the series
(\ref{eq:4.11}) turn out to be
\begin{equation}
{\cal A}_{i a} \stackrel{\kappa \to \infty}{\sim}
 \underbrace{i \, \sqrt{\frac{\Lambda}{3}} \, e_{i a}}_{{\cal O} (\kappa^{1/2})} +
 \underbrace{\begin{array}{c} { } \\[-.6 em] \omega_{i a} \\[-.6 em] { } \end{array}}_{
 {\cal O}(\kappa^0)} + \,
\underbrace{i \, \sqrt{\frac{3}{\Lambda}} 
 \left ( \frac{R}{4} \, e_{i a}  - e_{j a} \, {R^j}_i \right ) }_{
 {\cal O} ( \kappa^{-1/2} ) } + \,
 {\cal O} (\kappa^{-1}) \ .
\label{eq:4.12}
\end{equation}
To calculate the corresponding 
saddle-point contribution to the semiclassical Chern-Simons state via (\ref{eq:4.2.3})
we need the Gaussian prefactor and the exponent $F$ defined in 
(\ref{eq:4.2.2}), evaluated at the saddle-point ${\cal A}_{i a}$. The asymptotic form of
the Gaussian prefactor becomes in the limit $\kappa \to \infty$
\begin{equation}
\left | \,
\frac{\mu \, \delta^2 {F}}{\delta {\cal A}_{i a} (x) \, \delta {\cal A}_{j b} (y)}
\, \right |^{\mbox{\footnotesize{$-\frac{1}{2}$}}} 
\stackrel{\kappa \to \infty}{\propto}
\mbox{\raisebox{+.3 ex}{\footnotesize{$[$}} 
\hspace{-1.8 ex} \raisebox{-.15 ex}{\large{h}} $\!\!$}^{-3/4} \ , 
\label{eq:4.13}
\end{equation}
with the abbreviation
\begin{equation}
\mbox{\raisebox{+.3 ex}{\footnotesize{$[$}} 
\hspace{-1.8 ex} \raisebox{-.15 ex}{\large{h}} $\!\!$}
:=\prod_{x \in {\cal M}_3} \, h(x) \ .
\label{eq:4.14}
\end{equation}
The exponent in (\ref{eq:4.2.3}) for $\kappa \to \infty$ can be expanded
as follows:
\begin{equation}
{F} \, \stackrel{\kappa \to \infty}{\sim} \frac{1}{\gamma \hbar \mu} \left [ \,
 i \, \sqrt{\frac{3}{\Lambda}}
 \int d^3 x\,\sqrt{h} \left ( \frac{4 \Lambda}{3} -R \right ) +
 \frac{3}{\Lambda} \, {\cal S}_{\rm CS} ( \omega_{i a} ) \, \right ] 
 + \, {\cal O} (\kappa^{-3/2}) \ .
\label{eq:4.15}
\end{equation}
\\
Here the contribution $\phi$ from the similarity transformation (\ref{eq:3.8})
has disappeared, because it precisely cancels with the contribution
$\omega_{i a}$ in the asymptotic series (\ref{eq:4.12}) of ${\cal A}_{i a}$. The
first term in (\ref{eq:4.15})
derives from the contributions of order $\kappa^{1/2}$ and
$\kappa^{-1/2}$
to the asymptotic series of ${\cal A}_{i a}$ given in (\ref{eq:4.12}). 
It defines a real action 
\begin{equation}
\label{eq:4.17a}
S = \pm \frac{1}{\gamma} \, \sqrt{\frac{3}{\Lambda}} \, \int d^3 x \, \sqrt{h} \left (
\frac{4 \Lambda}{3} - R \right ) \ ,
\end{equation}
\\
giving rise to a well-defined, semiclassical time evolution.
The term of order $\kappa^{-1}$ in the expansion (\ref{eq:4.12}), 
which was not given explicitly there, because it is rather lengthy,
determines the asymptotic form of the second term in (\ref{eq:4.15}), which
is real-valued and
therefore governs the asymptotic behavior of $| \Psi_{\rm CS} |^2$. 
Surprisingly, this contribution again turns out
to be a Chern-Simons functional, but with ${\cal A}_{i a}$ replaced by the real
Riemannian spin-connection $\omega_{i a}$. As one can check quite easily,
this functional ${\cal S}_ {\rm CS} [ \omega_{i a} ]$ has the interesting property that it
is also invariant under \emph{local} 
scale-transformations of the triad $e_{i a} \mapsto \exp [ \zeta (x) ] \, e_{i a}$.

Inserting the results (\ref{eq:4.13})
and (\ref{eq:4.15}) into (\ref{eq:4.2.3}), we find for the semiclassical
Chern-Simons state in the $\tilde{e}^i \,\!_{a}$-representation
\begin{equation}
\Psi_{\rm CS} 
\  
\begin{array}{c} 
\mbox{\scriptsize{$\kappa \to \infty$}} \\[-1 ex] 
\mbox{\raisebox{-.1 ex}{$\propto$}} \\[-1 ex] 
\mbox{\scriptsize{$\mu \to \infty$}} 
\end{array}
\ \ {\cal N} \cdot 
\mbox{\raisebox{+.3 ex}{\footnotesize{$[$}} 
\hspace{-1.8 ex} \raisebox{-.15 ex}{\large{h}} $\!\!$}^{-3/4} 
\cdot \exp
\left [
\pm \frac{1}{\gamma \hbar} \left ( i \, \sqrt{\frac{3}{\Lambda}} \int d^3 x \, \sqrt{h}
\left ( \frac{4 \Lambda}{3} - R \right ) + \frac{3}{\Lambda} \, {\cal S}_{\rm CS}
[ \omega_{i a} ] \right )
\right ] \ ,
\label{eq:4.16}
\end{equation}
\\
where the complex conjugate solution $\Psi_{\rm CS}^*$ is equally possible, if we choose
the second saddle-point solution in eqs.~(\ref{eq:4.10}), (\ref{eq:4.10+}). 
It is remarkable that this result is universal in the sense
that it does not depend on the topology of the spatial 3-manifold
${\cal M}_3$.


\subsubsection{Large gauge-transformations}
\label{section4.2.0}

An unsatisfactory feature of the asymptotic state (\ref{eq:4.16}) is the fact that its exponent
is \emph{not} invariant under large gauge-transformations with a 
nonvanishing winding number:
As is well-known \cite{Ash1,Wein2}, in general the Chern-Simons functional 
${\cal S}_{\rm CS} [ \omega_{i a} ]$ transforms inhomogeneously 
under local gauge-transformations of the triad, 
\begin{equation}
  \label{eq:4.0.1}
  e_{i a} \mapsto \Omega_{a b} \, e_{i b} \quad \Rightarrow \quad
  {\cal S}_{\rm CS} [ \omega_{i a} ] \mapsto {\cal S}_{\rm CS} [ \omega_{i a} ] +
  {\textstyle \frac{1}{6}} \, I ( \bbox{\Omega} ) \ ,
\end{equation}
\\
with $( \Omega_{a b} ) = \bbox{\Omega} \in O(3)$ being an arbitrary rotation matrix.
The quantity $I ( \bbox{\Omega} )$ occuring in (\ref{eq:4.0.1}) is
defined by 
\begin{equation}
  \label{eq:4.0.2}
  I ( \bbox{\Omega} ) := \int d^3 x \, \tilde{\varepsilon}^{i j k} \, \mbox{Tr} \left [
  \bbox{\Omega}^{\mbox{\scriptsize{T}}}  \partial_i \bbox{\Omega} \cdot 
  \bbox{\Omega}^{\mbox{\scriptsize{T}}}  \partial_j \bbox{\Omega} \cdot
  \bbox{\Omega}^{\mbox{\scriptsize{T}}}  \partial_k \bbox{\Omega} \right ] 
\end{equation}
and known as the Cartan-Maurer invariant \cite{Wein2}.
Its value is restricted to be of the form
\begin{equation}
  \label{eq:4.0.3}
  I ( \bbox{\Omega} ) =I_0 \cdot w ( \bbox{\Omega} ) \ ,
\end{equation}
where the winding number $w ( \bbox{\Omega} )$ is an integer, and $I_0$ is
a constant depending only on the topology of the 3-manifold ${\cal M}_3$. 

A consequence of eq.~(\ref{eq:4.0.1}) is that the 
asymptotic Chern-Simons state (\ref{eq:4.16}) will not be
invariant under general gauge-transformations of the triad,
at least as long as we make a trivial choice for the normalization factor ${\cal N}$ 
in (\ref{eq:4.16}). However, as we pointed out in section~\ref{section3}, the
factor ${\cal N}$ does not need to be \emph{completely} independent of the triad
- it is still allowed to depend on topological invariants, such as the
Cartan-Maurer invariant. This is why we are free to choose the 
normalization factor ${\cal N}$ according to
\begin{equation}
  \label{eq:4.0.4}
  {\cal N} \propto \, \exp \left [ \, \mp \frac{I ( \hat{\bbox{\Omega}} )}{2 \gamma  \hbar \Lambda}
  \, \right ] \ ,
\end{equation}
where $\hat{\bbox{\Omega}}$ is a special gauge-transformation rotating the triad $e_{i a}$ 
into a \emph{gauge-fixed} triad $g_{i a}$ of the 3-metric $h_{i j}=e_{i a}  e_{j a}$. Then
the requirement (\ref{eq:3.17+}) remains to be satisfied, 
and, in addition, the Chern-Simons state
(\ref{eq:4.16}) becomes invariant under arbitrary gauge-transformations of the triad
$e_{i a}$, since the inhomogeneous term in (\ref{eq:4.0.1}) is cancelled precisely
by a suitable contribution from the prefactor ${\cal N}$ according to (\ref{eq:4.0.4}). 
With our
special choice (\ref{eq:4.0.4}) of the normalization factor ${\cal N}$ 
we circumvent the definition of the so-called ``$\Theta$-angle'', which can be 
introduced alternatively to solve the problem associated with
large gauge-transformations \cite{Ash1,Wein2}.
As a special, gauge-fixed triad $g_{i a}$ in the definition of $\hat{\bbox{\Omega}}$ may
serve the ``Einstein-triad'' that can be constructed by solving the eigenvalue problem
of the 3-dimensional Einstein-tensor ${G^i}_j$:\footnote{
Here a bar over an index indicates that \emph{no} summation with respect to this 
index should be performed.} 
\begin{equation}
  \label{eq:4.0.5}
  {G^i}_j \, g^j \,\!_a = \lambda_{\bar{a}} \, g^i \,\!_{\bar{a}} \qquad , 
 \qquad g^i \,\!_a \, g_{i b} = \delta_{a b} \ .
\end{equation}


\subsubsection{Restriction to Bianchi-type homogeneous 3-manifolds}
\label{section4.2.1}

It is very instructive to specialize the asymptotic state
(\ref{eq:4.16}) to the case of spatially homogeneous 3-manifolds. For homogeneous
manifolds of one of the nine Bianchi types, the 3-metric
can be expressed in terms of \emph{invariant} triad 1-forms 
$\bbox{e}_a =\bbox{\imath}_a = \imath_{i a} \, d x^i$ as (cf.~\cite{Ryan1})
\begin{equation}
 \bbox{h} = \bbox{\imath}_a \otimes \bbox{\imath}_a \qquad , \qquad
d \, \bbox{\imath}_a =
 - {\textstyle \frac{1}{2}} \, m_{b a} \, \varepsilon_{b c d} \, \bbox{\imath}_c \wedge
 \bbox{\imath}_d\,,
\label{eq:4.16a}
\end{equation}
with a \emph{spatially constant} structure matrix $\bbox{m} = ( m_{a b} )$. We should
restrict ourselves  to compactified, homogeneous 3-manifolds, 
such that the volume
\begin{equation}
V = {\textstyle \frac{1}{6}} \ \varepsilon_{a b c} \int 
\bbox{\imath}_a \wedge \bbox{\imath}_b \wedge \bbox{\imath}_c
\label{eq:4.16b}
\end{equation}
is finite. If we introduce the scale-invariant structure matrix $\bbox{M}$ as
\begin{equation}
 \bbox{M}= a_{\rm cos} \cdot \bbox{m} \ ,
\label{eq:4.16c}
\end{equation}
the asymptotic Chern-Simons state (\ref{eq:4.16}) takes the following value for 
Bianchi-type homogeneous 3-manifolds:
\begin{eqnarray}
\Psi_{\rm CS} 
\  
\begin{array}{c} 
\mbox{\scriptsize{$\kappa \to \infty$}} \\[-1 ex] 
\mbox{\raisebox{-.1 ex}{$\propto$}} \\[-1 ex] 
\mbox{\scriptsize{$\mu \to \infty$}} 
\end{array}
&& {\cal N} \cdot 
\mbox{\raisebox{+.3 ex}{\footnotesize{$[$}} 
\hspace{-1.8 ex} \raisebox{-.15 ex}{\large{h}} $\!\!$}^{-3/4} 
\cdot \exp
\Bigg [
\pm \mu \Bigg (4 i \, \sqrt{\kappa} - \frac{i}{\sqrt{\kappa}} \, 
\Bigl [ \,
\mbox{Tr} \, \bbox{M}^2 -2 \, \mbox{Tr} \, \bbox{M}^{\mbox{\scriptsize{T}}} \bbox{M} +
{\textstyle \frac{1}{2}} \, \mbox{Tr}^2 \, \bbox{M}
\, \Bigr ] \nonumber\\
&& \qquad \ 
- \frac{1}{\kappa} \Bigl [ \,
\mbox{Tr} \, \bbox{M}^2 \bbox{M}^{\mbox{\scriptsize{T}}} - {\textstyle \frac{1}{6}} \,
\mbox{Tr} \, \bbox{M}
\left ( \mbox{Tr} \, \bbox{M}^2 + 2 \, \mbox{Tr} \, \bbox{M}^{\mbox{\scriptsize{T}}} \bbox{M}
\right ) + 2 \, \det \bbox{M}
\Bigr ]
\Bigg )
\Bigg ] \ .
\label{eq:4.16e}
\end{eqnarray}
\\
For homogeneous manifolds of Bianchi-type~IX, the determinant of the matrix 
$\bbox{M}$ is given by $\det \bbox{M}=8 \, {\cal V}$, where ${\cal V}$ is the dimensionless,
invariant volume of the unit 3-sphere, so the matrix $\bbox{M}$ may be parametrized by 
a diagonal, traceless matrix $\bbox{\beta}$ via
\begin{equation}
\bbox{M}= 2 \, \sqrt[3]{\cal V} \, e^{\ 2 \bbox{\beta}} \ .
\label{eq:4.16f}
\end{equation}
Using the identity
\begin{equation}
\mbox{Tr} \, e^{2 \bbox{\beta}} \cdot \mbox{Tr} \, e^{4 \bbox{\beta}}
=
\mbox{Tr} \, e^{6 \bbox{\beta}} +
 \mbox{Tr} \, e^{2 \bbox{\beta}} \cdot \mbox{Tr} \, e^{-2 \bbox{\beta}} - 3 \ ,
\label{eq:4.16g}
\end{equation}
and introducing the rescaled parameter $\kappa' := {\cal V}^{-2/3} \, \kappa /4$, we find for
Bianchi-type~IX homogeneous 3-manifolds:
\begin{eqnarray}
\Psi_{\rm CS} 
\  
\begin{array}{c} 
\mbox{\scriptsize{$\kappa \to \infty$}} \\[-1 ex] 
\mbox{\raisebox{-.1 ex}{$\propto$}} \\[-1 ex] 
\mbox{\scriptsize{$\mu \to \infty$}} 
\end{array}
&& {\cal N} \cdot 
\mbox{\raisebox{+.3 ex}{\footnotesize{$[$}} 
\hspace{-1.8 ex} \raisebox{-.15 ex}{\large{h}} $\!\!$}^{-3/4} 
\cdot \exp
\Bigg [
\pm \frac{24 {\cal V}}{\gamma \hbar \Lambda} \Bigg ( 4 i \, \sqrt{\kappa'}^{\, 3} -
 i \, \sqrt{\kappa'} \,
\Bigl [
 \mbox{Tr} \, e^{-2 \bbox{\beta}} - {\textstyle \frac{1}{2}} \, \mbox{Tr} \, e^{4 \bbox{\beta}}
\Bigr ] \nonumber \\
&& \qquad \qquad \qquad \ \ \ - {\textstyle \frac{1}{2}} \, 
\Bigl [ \mbox{Tr} \,  e^{6 \bbox{\beta}} - \mbox{Tr} \, e^{2\bbox{\beta}} \cdot
\mbox{Tr} \, e^{-2 \bbox{\beta}} + 7 \Bigr ]
\Bigg )
\Bigg ] \ .
\label{eq:4.16h}
\end{eqnarray}
\\
Thus, up to a quantum correction in the Gaussian prefactor, we reproduce
exactly the result obtained earlier within the framework of the homogeneous
Bianchi~IX model in \cite{Pat1} (cf.~eq.~(5.18) there). 
To compare the results explicitly, we have to
identify $\kappa'$ with the parameter $\kappa$ in \cite{Pat1}, and to set 
$\gamma=16 \, \pi , {\cal V}=4 \, \pi^2$.

In the case of flat 3-metrics, which are of Bianchi-type~I, the structure matrix 
$\bbox{M}$ turns out to vanish, and (\ref{eq:4.16e}) reduces to 
\begin{equation}
\Psi_{\rm CS} 
\  
\begin{array}{c} 
\mbox{\scriptsize{$\kappa \to \infty$}} \\[-1 ex] 
\mbox{\raisebox{-.1 ex}{$\propto$}} \\[-1 ex] 
\mbox{\scriptsize{$\mu \to \infty$}} 
\end{array}
{\cal N} \cdot 
\mbox{\raisebox{+.3 ex}{\footnotesize{$[$}} 
\hspace{-1.8 ex} \raisebox{-.15 ex}{\large{h}} $\!\!$}^{-3/4} 
\cdot \exp
\left [ \,
\pm 4 i \, \mu \, \sqrt{\kappa}  \, \right ]
=
{\cal N} \cdot 
\mbox{\raisebox{+.3 ex}{\footnotesize{$[$}} 
\hspace{-1.8 ex} \raisebox{-.15 ex}{\large{h}} $\!\!$}^{-3/4} 
\cdot \exp
\left [ \, \pm \frac{4 i}{\gamma \hbar} \, 
 \sqrt{\frac{\Lambda}{3}} \int d^3 x \, \sqrt{h} \, \right ] \ ,
\label{eq:4.16i}
\end{equation}
\\
a result, which also follows directly from (\ref{eq:4.16}) by setting
$R=0$, ${\cal S}_{\rm CS } [ \omega_{i a} ] =0$.


\subsubsection{Semiclassical 4-geometries}
\label{section4.2.2}

Let us now ask for the semiclassical trajectories and the corresponding
semiclassical 4-geometries, which are described by the state
(\ref{eq:4.16}) in the limit $\mu \to \infty, \kappa \to \infty$, i.e. in the limit of
large scale-parameters $a_{\rm cos} \gg a_{\rm Pl}, a_{\Lambda}$. Choosing the Lagrangian
multipliers trivially as $N=1, N^i=0, \Omega_a=0$ in (\ref{eq:2.15a}), we find
\begin{equation}
\dot{\tilde{e}} \,\!^i \,\!_{a}
=
- \left \{ H \, , \, \tilde{e}^i \,\!_{a} \right \}
=
\pm i \, \tilde{\varepsilon}^{i j k} \, {\cal D}_j e_{k a}
=
- \frac{\gamma}{2} \, \tilde{\varepsilon}^{i j k} \,\varepsilon_{a b c} \, p_{j b} \, e_{k c} \ ,
\label{eq:4.17}
\end{equation}
where the dot denotes a derivative with respect to the classical 
ADM time-variable $t$ introduced in section~\ref{section2}.  
The semiclassical momentum $p_{i a}$ is given in terms of the action (\ref{eq:4.17a})
of the wavefunction (\ref{eq:4.16}) by
\begin{equation}
p_{i a} = \frac{\delta S}{\delta \tilde{e}^i \,\!_a} \ ,
\label{eq:4.18}
\end{equation}
or can equivalently be extracted from the asymptotic saddle-point ${\cal A}_{i a}$ 
according to
(\ref{eq:4.12}) in connection with (\ref{eq:2.11}):
\begin{equation}
p_{i a} \stackrel{\kappa \to \infty}{\sim} 
\pm \frac{2}{\gamma}
\left [
\sqrt{\frac{\Lambda}{3}} \, e_{i a} + \sqrt{\frac{3}{\Lambda}} \, 
\left ( \frac{R}{4} \, e_{i a} - e_{j a} \, {R^j}_i \right )
\right ] \ .
\label{eq:4.19}
\end{equation}
\\
Thus, for large scale-parameters $a_{\rm cos}$ the classical evolution
of the triad $\tilde{e}^i \,\!_{a}$ is determined by the equation
\begin{equation}
\mp \, \dot{\tilde{e}} \,\!^i \,\!_{a} \stackrel{^{\mbox{\scriptsize{$a_{\rm cos} \to \infty$}}}}{\sim}
2 \, \sqrt{\frac{\Lambda}{3}} \, \tilde{e}^i \,\!_{a} +
 \sqrt{\frac{3}{\Lambda}} \, \tilde{e}^j \,\!_{a} {G^i}_j \ ,
\label{eq:4.20}
\end{equation}
\\
which describes a deSitter-like time-evolution in leading order $a_{\rm cos}$,
\begin{equation}
\tilde{e}^i \,\!_{a} (x,t) \stackrel{^{\mbox{\scriptsize{$a_{\rm cos} \to \infty$}}}}{\sim}
 \tilde{e}^i \!\,_{a , \infty} (x)
\cdot \exp \left [ \mp \, 2 \, \sqrt{\frac{\Lambda}{3}} \cdot t \right ] \ ,
\label{eq:4.21}
\end{equation}
\\
with corrections described by the second term of (\ref{eq:4.20}) containing
the 3-dimensional Einstein-tensor ${G^i}_j$. 

The figure $1$ shows an embedding
of the asymptotic 4-geometry (\ref{eq:4.21}) into a flat Minkowski space, where the
time direction has been chosen according to the lower sign in eq.~(\ref{eq:4.21}). 
As is well-known for inflationary models like the one discussed within this paper,
the spatial, Riemannian 3-manifolds $( {\cal M}_3 , \bbox{h} ) (t)$ tend to homogenize
in the course of time $t$.


\begin{center}
\setlength{\unitlength}{.8 \textwidth}
\begin{picture}(1,.92)(0,0)

   \put(.635,.225){$( {\cal M}_3  , \bbox{h}  ) (t_0)$}
   \put(.63,.45){$t$}
   \put(.82,.69){$t \to \infty$}
   \put(.32,.05){$a$}
   \put(.32,.18){$a$}
   \put(.73,.52){$( {\cal M}_4 , \bbox{g} )$}
   \put(0.05,0){
   \leavevmode
   \epsfxsize= \unitlength
   \epsffile{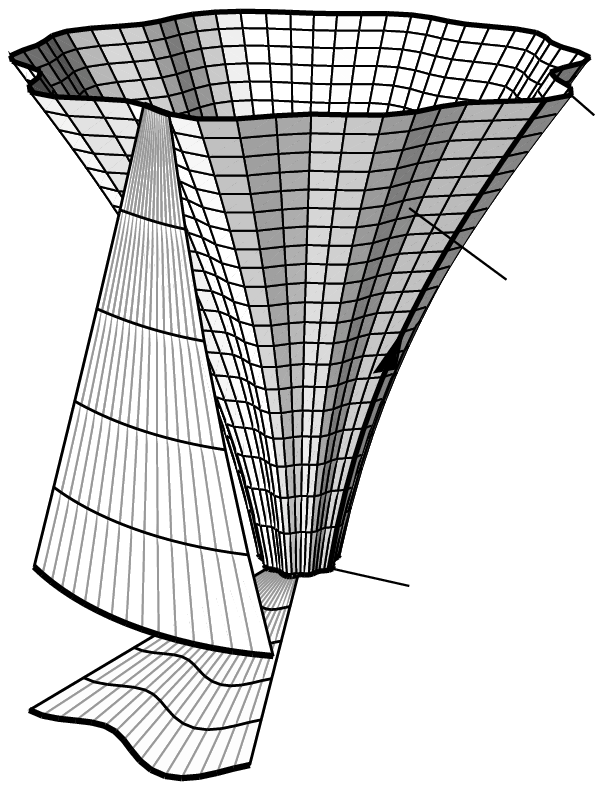}}
\end{picture}
\end{center}

\begin{minipage}{.9 \textwidth}
\small{FIG.~$1$.}
\footnotesize{Geometrical illustration of the generalized deSitter-4-geometry 
(\ref{eq:4.21}).
The spatial 3-manifolds $( {\cal M}_3 , \bbox{h} ) (t)$ are
represented by 1-dimensional curves, possible inhomogenieties are indicated by small
deformations of these curves. The resulting space-time 4-manifold 
$( {\cal M}_4 , \bbox{g} )$
according to (\ref{eq:4.21}) then corresponds to a 2-dimensional, Lorentzian manifold,
which has been embedded into a flat, 3-dimensional Minkowski space. Portions of the
marginal spatial 3-manifolds, which are of the \emph{same} length-scale $a$,
have been magnified to illustrate the increase in homogeneity in the course of evolution.}
\end{minipage}


\subsection{The semiclassical vacuum limit $\mu \to \infty, \kappa \to 0$}
\label{section4.3}

Apart from the limit $\kappa \to \infty$ there exists another asymptotic regime, where
an analytical treatment of the semiclassical saddle-point equations (\ref{eq:4.9+})
is tractable, namely the limit $\kappa \to 0$. 
By virtue of the relationships (\ref{eq:4.1.4}), a discussion of the 
Chern-Simons state
(\ref{eq:4.2.1}) in the limit $\mu \to \infty , \kappa \to 0$ corresponds to an investigation
of the asymptotic regime $a_{\Lambda} \gg a_{\rm cos} \gg a_{\rm Pl}$. This limit may
be realized by considering the special case of a vanishing cosmological constant
$\Lambda \to 0$ within the semiclassical limit, what will be called the
semiclassical vacuum limit for short. 

To find solutions of eqs.~(\ref{eq:4.9+}) in the limit $\kappa \to 0$ we proceed
analogously to section~\ref{section4.2}, and try a power series ansatz of the form 
\begin{equation}
{\cal A}_{i a} \stackrel{\kappa \to 0}{\sim} \sum_{n=0}^{\infty} C_{i a}^{(n)} \, \kappa^n \ .
\label{eq:4.30.1}
\end{equation}
Then we find in the lowest order of $\kappa$
\begin{equation}
  \label{eq:4.30.2}
  \tilde{\varepsilon}^{i j k} \, \left (
  \partial_j C_{k a}^{(0)} + {\textstyle \frac{1}{2}} \, \varepsilon_{a b c} \, C_{j b}^{(0)} \,
  C_{k c}^{(0)} \right ) = 0 \ ,
\end{equation}
\\
i.e. $C_{i a}^{(0)}$ has to be a \emph{flat} gauge-field, which is of the general form
\begin{equation}
  \label{eq:4.30.3}
  C_{i a}^{(0)} = - {\textstyle \frac{1}{2}} \, \varepsilon_{a b c} \, \Omega_{d b} \, \partial_i
  \, \Omega_{d c} \qquad \mbox{with} \qquad \bbox{\Omega} \in O(3) \ .
\end{equation}
\\
The matrix $\bbox{\Omega} (x)$ is a free integration field, as long as we restrict
ourselves to the leading order ${\cal O} (\kappa^0)$ of the saddle-point equations
(\ref{eq:4.9+}). However, in the next to leading order ${\cal O} (\kappa^1)$, we
find the equations
\begin{equation}
\label{eq:4.30.4}
\tilde{\varepsilon}^{i j k} \, {\cal D}_j^{(0)} \, C_{k a}^{(1)} :=
\tilde{\varepsilon}^{i j k} \left (
\partial_j C_{k a}^{(1)} + \varepsilon_{a b c} \, C_{j b}^{(0)} \, C_{k c}^{(1)} \right ) 
\stackrel {!}{=}  -
\, \tilde{e}' \,\!^i \,\!_a \ ,
\end{equation}
\\
which imply additional restrictions for the coefficients $C_{i a}^{(0)}$, and thus for
the matrix $\bbox{\Omega}$ in (\ref{eq:4.30.3}). These \emph{integrability conditions} for
the equations (\ref{eq:4.30.4}) can be obtained by operating on (\ref{eq:4.30.4}) with
${\cal D}_{i}^{(0)}$ from the left: Then the left hand side becomes proportional to the
curvature of $C_{i a}^{(0)}$, which vanishes by virtue of eq.~(\ref{eq:4.30.2}), and
a multiplication of the resulting equations with $a_{\rm cos}^2$ yields
\begin{equation}
  \label{eq:4.30.5}
  {\cal D}_i^{(0)} \, \tilde{e}^i \,\!_a \equiv \partial_i \tilde{e}^i \,\!_a + \varepsilon_{a b c} \,
  C_{i b}^{(0)} \, \tilde{e}^i \,\!_c \stackrel{!}{=} 0 \ .  
\end{equation}
If we insert the general solution (\ref{eq:4.30.3}) into (\ref{eq:4.30.5}), we arrive at
the three integrability conditions
\begin{equation}
  \label{eq:4.30.6}
  \partial_i \left ( \Omega_{a b} \, \tilde{e}^i \,\!_b \right ) \stackrel{!}{=} 0 \ ,
\end{equation}
which fix the integration field $\bbox{\Omega} (x)$ in (\ref{eq:4.30.3}). Moreover,
the special triad fields
\begin{equation}
  \label{eq:4.30.7}
  \tilde{d}^{\, i} \,\!_a := \Omega_{a b} \, \tilde{e}^i \,\!_b
\end{equation}
with $\bbox{\Omega}$ chosen according to (\ref{eq:4.30.6}) turn out to have the
geometrically interesting property of being \emph{divergence-free}. Therefore, we may
use the different possible divergence-free triads $\tilde{d}^{\, i} \,\!_a$ of a 
given Riemannian manifold
$( {\cal M}_3 , \bbox{h} )$ to parameterize the saddle-points ${\cal A}_{i a}$ 
in the limit $\kappa \to 0$ via (\ref{eq:4.30.7}) and (\ref{eq:4.30.3}).

For a given divergence-free triad $\tilde{d}^{\, i} \,\!_a$,
which characterizes uniquely one saddle-point solution ${\cal A}_{i a}$ in the limit
$\kappa \to 0$, 
we now wish to calculate the corresponding saddle-point contribution
(\ref{eq:4.2.3}) to the Chern-Simons state (\ref{eq:4.2.1}) in the limit 
$\mu \to \infty, \kappa \to 0$.  We first expand the exponent $F$ defined in (\ref{eq:4.2.2})
for $\kappa \to 0$, and find, in particular, that the Chern-Simons functional
${\cal S}_{\rm CS} [ {\cal A}_{i a} ]$ is given by
\begin{equation}
  \label{eq:4.30.8}
  {\cal S}_{\rm CS} [ {\cal A}_{i a} ] \stackrel{\kappa \to 0}{\sim} 
  {\textstyle \frac{1}{6}} \, I ( \bbox{\Omega} ) + {\cal O} ( \kappa^2 ) \ .
\end{equation}
Here $\bbox{\Omega}$ is the special rotation matrix 
defined in (\ref{eq:4.30.6}), connecting the
given divergence-free triad $\tilde{d}^{\, i} \,\!_a$ with an arbitrary triad $\tilde{e}^i \,\!_a$,
for which we want to evaluate $\Psi_{\rm CS} [ \tilde{e}^i \,\!_a ]$. In (\ref{eq:4.30.8})
a contribution of order ${\cal O} ( \kappa^1 )$ is missing, since this term becomes
proportional to the curvature of the \emph{flat} gauge-field $C_{i a}^{(0)}$. Using
eq.~(\ref{eq:4.30.8}), the exponent $F$ of the semiclassical Chern-Simons state
takes the following form in the limit $\kappa \to 0$:
\begin{equation}
  \label{eq:4.30.9}
  F \stackrel{\kappa \to 0}{\sim} \frac{I ( \bbox{\Omega} )}{6 \, \kappa} +
  \int d^3 x \, \tilde{\varepsilon}^{i j k} \, d'_{i a} \, \partial_j \, d'_{k a} +
  {\cal O} ( \kappa ) \ .
\end{equation}
\\
The Cartan-Maurer invariant $I ( \bbox{\Omega} )$ in (\ref{eq:4.30.9}) can be
contracted with the Cartan-Maurer invariant  $I ( \hat{\bbox{\Omega}} )$ in the
definition (\ref{eq:4.0.4}) of the normalization factor ${\cal N}$ to give
\begin{equation}
  \label{eq:4.30.10}
  I ( \bbox{\Omega} ) - I ( \hat{\bbox{\Omega}} ) \equiv I  ( \bbox{\Omega} \cdot
  \hat{\bbox{\Omega}} \,\!^{\mbox{\scriptsize T}}  ) =:
  I_0 \cdot \hat w [ d_{i a} ] \ .
\end{equation}
Here $\hat w [ d_{i a} ]$ denotes the winding number of the divergence-free
triad $d_{i a}$ with respect to the Einstein-triad $g_{i a}$ defined in
(\ref{eq:4.0.5}), which is a functional of $d_{i a}$ only: For a given divergence-free
triad $d_{i a}$ we know the 3-metric $h_{i j} = d_{i a} \, d_{j a}$, and therefore the
Einstein-triad $g_{i a}$.

Inserting the results (\ref{eq:4.30.10}), (\ref{eq:4.30.9}) into (\ref{eq:4.2.3}), we find
the following saddle-point contribution to the Chern-Simons state (\ref{eq:4.2.1})
in the limit $\mu \to \infty, \kappa \to 0$ 
 \begin{equation}
\lim_{\kappa \to 0} \, \Psi_{\rm CS}
=: \Psi_{\rm vac} 
\stackrel{\mu \to \infty}{\propto}
\exp
\left [ \pm 
\frac{1}{\gamma \hbar} \left (
\frac{I_0 \, \hat w [ d_{i a} ]}{2 \,\Lambda} + \int d^3 x \,\tilde{\varepsilon}^{i j k}
\, d_{i a} \, \partial_j \, d_{k a} \right ) \right ]  \ ,
\label{eq:4.31}
\end{equation}
\\
where the Gaussian prefactor, which contains a complicated, 
non-local functional determinant, has been hidden in the proportionality sign.

From the result (\ref{eq:4.31}) we can see the gauge-invariance of the 
semiclassical vaccum state $\Psi_{\rm vac}$, since this state does not depend
explicitly on the triad $\tilde{e}^i \,\!_a$, but only on the 3-metric $h_{i j}=e_{i a} \, e_{j a}$,
to which we have chosen a \emph{fixed} divergence-free triad $\tilde{d}^{\, i} \,\!_a$.
It is remarkable that for the one unique choice (\ref{eq:4.0.4}) of the prefactor ${\cal N}$
gauge-invariance, even under large gauge-transformations, can be achieved in  
both of the two quite different limits $\kappa \to \infty$ and $\kappa \to 0$.  

The \emph{existence} of divergence-free triads to a given 3-metric $h_{i j}$ is
discussed in appendix~\ref{appA.2}. There, we also argue that in general there will
even exist different, topologically inequivalent divergence-free triads, giving
rise to linearly independent semiclassical vacuum states via (\ref{eq:4.31}).


\subsubsection{Restriction to Bianchi-type~A homogeneous 3-manifolds}
\label{section4.3.1}

We now wish to evaluate the semiclassical vacuum state (\ref{eq:4.31}) for the special
case of Bianchi-type homogeneous 3-manifolds. For such manifolds, it follows directly
from (\ref{eq:4.16a}) that the divergence of the \emph{invariant} triad 
$\vec{\imath}_a = \imath^{\, i} \,\!_a \, \partial_i$ can
be expressed in terms of the structure matrix $\bbox{m}$ as
\begin{equation}
  \label{eq:4.32}
  \vec{\nabla} \cdot \vec{\imath}_a = \frac{1}{\sqrt{h}} \, 
  \partial_i \, \tilde{\imath}^{\, i} \,\!_a
 = \varepsilon_{a b c} \, m_{b c} \ .
\end{equation}
Consequently, the invariant triad $\vec{\imath}_a$ of Bianchi-type homogeneous
3-manifolds is divergence-free, if, and only if the structure matrix $\bbox{m}$ is
\emph{symmetric}, i.e. if the 3-manifold is of Bianchi-type~A. If we restrict ourselves
to this special class of manifolds in the following, at least \emph{one} divergence-free
triad $\vec{d}^{\, (0)}_a = \vec{\imath}_a$ is known, and we can calculate the corresponding
value of the semiclassical vacuum state (\ref{eq:4.31}):
\begin{equation}
  \label{eq:4.32.1}
  \Psi^{(0)}_{\rm vac} 
  \stackrel{\mu \to \infty}{\propto} 
  \exp \left [ \mp \, \frac{V}{\gamma \hbar} \, \mbox{Tr} \, \bbox{m}
  \right ] \ .
\end{equation}
\\
Here we made use of the fact that for 3-manifolds of Bianchi-type~A the invariant
triad $\vec{\imath}_a$ and the Einstein-triad $\vec{g}_a$ differ only by a 
\emph{spatially constant} rotation $\hat{\bbox{\Omega}}$, implying a vanishing winding number
$\hat{w} [ \imath_{i a} ] = 0$ in (\ref{eq:4.31}).
A further specialization of the result (\ref{eq:4.32.1}) to Bianchi-type~IX homogeneous
manifolds gives
\begin{equation}
  \label{eq:4.32.2}
  \Psi^{(0)}_{\rm vac} \stackrel{\mu \to \infty}{\propto}  
  \exp \left [ \mp \, \frac{2 {\cal V}}{\gamma \hbar} \left (
  a_1^2 + a_2^2 + a_3^2 \right ) \right ] \ ,
\end{equation}
\\
where we have introduced the three scale parameters ${a}_{b}$ via
\begin{equation}
\label{eq:4.32.3}
\bbox{m} =:
2 \, \mbox{diag} \left [ \frac{a_1}{a_2 \, a_3} , \frac{a_2}{a_3 \, a_1} , \frac{a_3}{a_1 \, a_2}
\right  ] \qquad  \Rightarrow \qquad V = {\cal V} \, a_1 \, a_2 \, a_3 \ ,
\end{equation}
\\
with the same, dimensionless volume ${\cal V}$ of the unit 3-sphere that already occured in
section~\ref{section4.2.1}. The saddle-point value (\ref{eq:4.32.2}) corresponds to the
``wormhole-state'' of the Bianchi~IX model \cite{Pat1,Pat3}. 
Within the framework of the homogeneous Bianchi~IX model,
four further semiclassical vacuum states are known, 
which, in the inhomogeneous approach of the present paper, correspond to nontrivial
divergence-free triads of Bianchi-type~IX manifolds via (\ref{eq:4.31}). These
topologically nontrivial divergence-free triads of Bianchi-type~IX metrics
and the resulting values of the semiclassical vacuum
state (\ref{eq:4.31}) will be discussed separately in appendix~\ref{appB}.

As a further restriction of the state (\ref{eq:4.32.1}) one may consider again the
case of flat Bianchi-type~I manifolds, where the structure matrix $\bbox{m}$, and
therefore the exponent of (\ref{eq:4.32.1}), vanishes. Thus, for flat
3-manifolds the behavior of the semiclassical vacuum state is governed by the 
Gaussian prefactor, which we do not know explicitly.


\subsubsection{Semiclassical 4-geometries}
\label{section4.3.2}

The semiclassical trajectories and the associated 4-geometries, which
are generated by the state  (\ref{eq:4.31}) in the limit $\kappa \to 0, \mu \to \infty$,
can be calculated by solving the evolution equations (\ref{eq:4.17}) with the 
flat, semiclassical
spin-connection ${\cal A}_{i a}$ derived in section~\ref{section4.3}. However, in contrast
to the limit $\kappa \to \infty$ discussed in section~\ref{section4.2.2}, we here arrive
at \emph{imaginary} evolution equations, since the semiclassical action of the
wavefunctional $\Psi_{\rm vac}$ according to (\ref{eq:4.31}) is purely imaginary. 
Following Hawking \cite{Haw1}, a geometrical interpretation may still be given in terms
of an imaginary time variable $\tau := i \, t$, converting the Lorentzian signature
of the 4-dimensional space-time into a positive, Euclidian signature. Then the 
semiclassical evolution equations can conveniently be expressed
in terms of the divergence-free triad $d_{i a}$, which characterizes the flat
Ashtekar spin-connection ${\cal A}_{i a}$ in the limit $\kappa \to 0$:
\begin{equation}
  \label{eq:4.40.0}
  \frac{d}{d \tau} \, \tilde{d}^{\, i} \,\!_{a} = \pm \, \tilde{\varepsilon}^{i j k} \, \partial_j \, d_{k a} 
  \qquad \Leftrightarrow \qquad \frac{d}{d \tau} \, d_{i a} = \mp \, \omega_{i a} \ .
\end{equation}
Here $\omega_{i a}$ in the second equation is the Riemannian spin-connection of the
divergence-free triad $d_{i a}$. Obviously, the gauge-condition 
$\partial_i \, \tilde{d}^{\, i} \,\!_a =0$ remains preserved in the course of evolution, 
as it must be the case. 

Stationary solutions of eqs.~(\ref{eq:4.40.0}) are given by $\omega_{i a}=0$, i.e. \emph{flat}
3-manifolds $( {\cal M}_3 , \bbox{h} )$. With our trivial choice of the Lagrangian
multipliers $N=1, N^i=0$, these correspond to locally flat, positive definite semiclassical
space-time manifolds $( {\cal M}_4 , \bbox{g} )$. Further solutions of (\ref{eq:4.40.0}) can
be constructed with help of the scaling ansatz
\begin{equation}
  \label{eq:4.41}
  d_{i a} ( x , \tau) = \mp \, \tau \cdot {d'}_{i a} (x) \ ,
\end{equation}
which implies ${d'}_{i a} (x) = \omega_{i a} (x)$, and therefore a simple form for 
the Ricci-tensor of the spatial 3-manifold:
\begin{equation}
  \label{eq:4.42}
  {R^i}_j = \frac{2}{\tau^2} \, \delta^i_j \ .
\end{equation}
Consequently, the spatial manifold has to be a 3-sphere with radius $\tau$, and
the 4-dimensional line element becomes
\begin{equation}
  \label{eq:4.43}
  d s^2 = d \tau^2 + \tau^2 \, d \Omega^2_3 \ ,
\end{equation}
with $d \Omega^2_3$ being the line element of the unit 3-sphere. As for the
stationary solutions mentioned above, the line element (\ref{eq:4.43}) describes a 
locally flat, positive definite 4-manifold.

Because of the nonlinearity of the evolution equations (\ref{eq:4.40.0}),  the general
behavior of the solutions is quite complicated and cannot be discussed here. However,
a complete discussion of the possible semiclassical trajectories can be given within the
narrow class of Bianchi-type~IX homogeneous 3-manifolds, cf.~\cite{Pat1}. There it turns
out, that the semiclassical evolution governed by the invariant, divergence-free
triad $\vec{d}^{\, (0)}_a = \vec{\imath}_a$, which corresponds to the ``wormhole-state''
(\ref{eq:4.32.2}) via (\ref{eq:4.31}), 
gives rise to asymptotically \emph{flat} 4-geometries in the limit of large
scale parameters $a_{\rm cos}$. Moreover, a second divergence-free triad of 
these Bianchi-type~IX homogeneous 3-manifolds, 
which is given in appendix~\ref{appB}, is known
to evolve in such a way, that \emph{compact, regular} 4-manifolds are approached
in the limit of vanishing scale parameter $a_{\rm cos}$.\footnote{The  
semiclassical vacuum state, corresponding to this second divergence-free triad
via (\ref{eq:4.31}), is the ``no-boundary-state'' of the Bianchi~IX model.}

One may now ask, if such a universal behavior of the semiclassical trajectories, that
can be found within the Bianchi~IX model, carries over to the inhomogeneous
case. Unfortunately, this does not seem to be the case: In appendix~\ref{appC}
we explicitly solve the evolution equations (\ref{eq:4.40.0}) for a
particular class of initial 3-manifolds, and find, that these solutions
neither satisfy the condition of asymptotical
flatness in the limit $a_{\rm cos}\ \to \infty$, nor the ``no-boundary'' proposal suggested
by Hartle and Hawking \cite{Haw1,Haw2,Haw3}. Thus we conclude that, in the
inhomogeneous case, the
semiclassical vacuum state given in (\ref{eq:4.31}) will in general \emph{not} be subject to
any specific boundary condition, like the ``no-boundary'' proposal or the
condition of asymptotical flatness.


\section{Non-normalizability of the Chern-Simons state in a physical
inner product}
\label{section5}

We now want to argue that the gravitational Chern-Simons state 
$\Psi_{\rm CS} [ \tilde{e}^i \,\!_a ]$ according to eq.~(\ref{eq:3.17}) 
does \emph{not} constitute a normalizable physical
state on the Hilbert space of quantum gravity. Therefore, we will derive a
physical inner product on the configuration space of real triads, 
which we want to be gauge-fixed with respect to the time-reparametrization 
invariance of general relativity. In this particular inner product, we then will 
try to calculate the corresponding norm
of the Chern-Simons state $\Psi_{\rm CS} [ \tilde{e}^i \,\!_a ]$. 

To derive a physical inner product
within the framework of the Faddeev-Popov calculus \cite{FadPop,Wood}, 
we first have to find
a kinematical inner product, denoted by $\langle \cdot |\cdot \rangle$ in the
following, with respect to which the quantum constraint operators $\tilde{{\cal H}}_0$,
$\tilde{{\cal H}}_i$ and $\tilde{{\cal J}}_a$ are formally hermitian. 
Since the \emph{complex} Hamiltonian constraint
operator $\tilde{{\cal H}}_0$ defined in eq.~(\ref{eq:2.13a}) cannot be hermitian
with respect to \emph{any} inner product on the configuration space, we replace
$\tilde{{\cal H}}_0$ by its real version $\tilde{{\cal H}}_0^{\rm ADM}$ given in
(\ref{eq:2.13}), with the factor ordering suggested there. With the help of the
commutators (\ref{eq:3.0.1})-(\ref{eq:3.0.6}) one can check quite easily that the algebra of
$\tilde{{\cal H}}_0^{\rm ADM}, \tilde{{\cal H}}_i$ and $\tilde{{\cal J}}_a$ still closes without
any quantum corrections. 
However, the
explicit commutators turn out to be much more complicated than
the corresponding commutators of $\tilde{{\cal H}}_0, \tilde{{\cal H}}_i, \tilde{{\cal J}}_a$ 
given in (\ref{eq:3.0.1})-(\ref{eq:3.0.6}), and will not be given here.

Since the quantum state $\Psi_{\rm CS}$ given in (\ref{eq:3.17}) is also
annihilated by the operator $\tilde{{\cal H}}_0^{\rm ADM}$, the substitution
$\tilde{{\cal H}}_0 \mapsto \tilde{{\cal H}}_0^{\rm ADM}$ 
has no negative consequences for the
theory, but the positive effect that we can now define a kinematical inner
product, with respect to which the operators $\tilde{{\cal H}}_0^{\rm ADM}, \tilde{{\cal H}}_i$
and $\tilde{{\cal J}}_a$ are hermitian. This product turns out to be
\begin{equation}
\langle \Psi | \Phi \rangle =
\int {\cal D}^9 \, [ e_{i a} ] \, \Psi^* [ e_{i a} ] \cdot \Phi [ e_{i a} ] \ ,
\label{eq:5.1}
\end{equation}
where the functional integral has to be performed over all real triads
$e_{i a} (x)$. While $\tilde{{\cal H}}_0^{\rm ADM}$ and $\tilde{{\cal J}}_a$ are formally hermitian
in the product (\ref{eq:5.1}), $\tilde{{\cal H}}_i$ is hermitian only if we
take a regularization of the theory, where terms containing the singular
object $(\partial_i \delta) (0)$ vanish.\footnote{Some authors argue that
this should be possible, cf.~Matschull \cite{Mat}.} 
If we can achieve this,
we have found a kinematical inner product on the configuration space of all
real triads $e_{i a} (x)$, and can continue with the Faddeev-Popov calculus
by choosing a gauge-condition $\tilde{\chi} [ e_{i a} ] = 0$ fixing the time-gauge. 
The corresponding physical inner product is then obtained as
\begin{equation}
\langle \! \langle \Psi | \!\!\; | \Phi \rangle \! \rangle_{\rm phys}
=
\langle \Psi  | \,  \delta [ \tilde{\chi} ] \cdot | J_{\rm H} |  \,  | \Phi \rangle \ ,
\label{eq:5.2}
\end{equation}
with the Faddeev-Popov functional determinant 
\begin{equation}
  \label{eq:5.2+}
  J_{\rm H}  := \det \left ( \, \frac{i}{\hbar}
 \left [ \, \tilde{{\cal H}}_0^{\rm ADM} (x) \, , \, \tilde{\chi} (y) \, \right ] \, \right )  \ .
\end{equation}

A rather natural way to fix the time-gauge is to consider 3-geometries with a
given volume-form $\sqrt{h (x)}$, for which there remain only two local degrees of freedom.
Therefore we assume
$\tilde{\upsilon} (x)$ to be a fixed, positive scalar density of weight $+1$
on the spatial manifold ${\cal M}_3$, normalized such that\footnote{
For example, the quantity $\tilde{\upsilon}$ may be chosen as the 
rescaled volume element 
of a maximally symmetric 3-metric on ${\cal M}_3$.} 
\begin{equation}
\int d^3 x \, \tilde{\upsilon} (x) \stackrel{!}{=} 1 \ .
\label{eq:5.3}
\end{equation}
Furthermore, let $a_{\chi}$ be an arbitrary, positive scale parameter. 
Then the gauge-condition
\begin{equation}
\tilde{\chi} := \sqrt{h(x)} -a_{\chi}^3 \, \tilde{\upsilon} (x) 
\stackrel{!}{=} 0 
\label{eq:5.4}
\end{equation}
is a diffeomorphism- and $SO(3)$-gauge-invariant equation fixing the volume-form
of the 3-metric. In particular, it follows from eq.~(\ref{eq:5.4}) that the length scale
$a_{\chi}$ and the cosmological scale $a_{\rm cos}$ introduced in (\ref{eq:4.1.2})
must be equal.
In the gauge (\ref{eq:5.4}), the physical norm associated with the inner product
(\ref{eq:5.2}) obviously depends on the scale parameter $a_{\chi}$
and the choice of $\tilde{\upsilon} (x)$, but we can consider the limit $a_{\chi} \to \infty$,
\begin{equation}
| \!\!\; | \Psi | \!\!\; |_\infty^2
:=
\lim_{a_{\chi} \to \infty} \langle \! \langle \Psi | \!\!\; | \Psi \rangle \! \rangle_{\rm phys} \ ,
\label{eq:5.5}
\end{equation}
which, in case of the Chern-Simons state $\Psi = \Psi_{\rm CS}$, 
will turn out to be independent of $\tilde{\upsilon} (x)$. For an
explicit calculation of (\ref{eq:5.5}), we need the Faddeev-Popov
commutator occuring in (\ref{eq:5.2+}), which turns out to be
\begin{equation}
\frac{i}{\hbar} \left [ \, \tilde{{\cal H}}_0^{\rm ADM} (x) \, , \, \tilde{\chi} (y) \, \right ]
=
\frac{\gamma}{4} \, \delta^3 (x-y) \, \tilde{\jmath} (x) \ ,
\label{eq:5.6}
\end{equation}
with
\begin{equation}
  \label{eq:5.7}
  \tilde{\jmath} (x) := \frac{ i \hbar}{2} \, \left [ e_{i a} (x) \, \frac{\delta}{\delta e_{i a} (x)} +
\frac{\delta}{\delta e_{i a} (x)} \, e_{i a} (x) \right ] \ .
\end{equation}
\\
The Faddeev-Popov functional determinant $J_{\rm H}$ according to
(\ref{eq:5.2+}) follows as
\begin{equation}
J_{\rm H} =
\prod_{x \in {\cal M}_3} \, \frac{\gamma}{4} \, \tilde{\jmath} (x) \ ,
\label{eq:5.8}
\end{equation}
which, acting on the wavefunctional $\Psi_{\rm CS}$, measures the space product
of the current $\tilde{\jmath} (x)$ of $\Psi_{\rm CS}$ in the $h(x)$-direction of superspace. 
Since we are dealing with the limit $a_{\chi} = a_{\rm cos} \to \infty$, 
the exact quantum state $\Psi_{\rm CS}$
given in (\ref{eq:3.17}) may be substituted by the asymptotic state (\ref{eq:4.16})
for explicit calculations.
Then the current of $\Psi_{\rm CS}$ in the $h(x)$-direction
turns out to have the same sign at each space-point for large scale parameters 
$a_{\chi} = a_{\rm cos}$,\footnote{
This property of $\Psi_{\rm CS}$ in the limit $a_{\rm cos} \to \infty$
reminds one of the Vilenkin proposal for the wavefunction of the Universe discussed
in \cite{Vil1,Vil3}.}
so we do \emph{not} need to take the
modulus of the Faddeev-Popov determinant in (\ref{eq:5.2}), as the general
calculus in \cite{FadPop} would prescribe. More explicitly, we find the result
\begin{equation}
J_{\rm H} \cdot \Psi_{\rm CS} \, \Big |_{\tilde{\chi} = 0} 
\stackrel{^{\mbox{\scriptsize{$a_{\chi} \to \infty$}}}}{\propto} 
\mbox{\raisebox{+.12 ex}{\footnotesize{$[$}} 
\hspace{-1.8 ex} \raisebox{-.35 ex}{\large{h}} $\!\!$}^{1/2} \cdot \Psi_{\rm CS} 
\, \Big |_{\tilde{\chi} = 0} \ \ ,
\label{eq:5.9}
\end{equation}
\\
where 
$\mbox{\raisebox{+.42 ex}{\footnotesize{$[$}} 
\hspace{-1.8 ex} \raisebox{-.05 ex}{\large{h}} $\!\!$}$ 
was defined in (\ref{eq:4.14}), so the physical norm
(\ref{eq:5.5}) becomes in the limit $a_{\chi} \to \infty$:
\begin{equation}
| \!\!\; | \Psi_{\rm CS} | \!\!\; |_\infty^2 \propto \int {\cal D}^9 [ e_{i a} ] \,
\mbox{\raisebox{+.12 ex}{\footnotesize{$[$}} 
\hspace{-1.8 ex} \raisebox{-.35 ex}{\large{h}} $\!\!$}^{1/2} \, | \Psi_{\rm CS} |^2 \, 
\delta [ \tilde{\chi} ] \ .
\label{eq:5.10}
\end{equation}
\\
If we now introduce the new integration variables $\sqrt{h}$, and eight
locally scale-invariant fields $\beta_\kappa$, the functional integral in
(\ref{eq:5.10}) becomes
\begin{eqnarray}
 | \!\!\; | \Psi_{\rm CS} | \!\!\; |_\infty^2
&\propto &
\int {\cal D} [ \sqrt{h} ] {\cal D}^8 [ \beta_\kappa ] \, w [ \beta_\kappa] \,
\mbox{\raisebox{+.12 ex}{\footnotesize{$[$}} 
\hspace{-1.8 ex} \raisebox{-.35 ex}{\large{h}} $\!\!$}^{3/2} 
\, | \Psi_{\rm CS} |^2 \, \delta \big [ \sqrt{h} - a_{\rm cos}^3 \, \tilde{\upsilon} \big ] 
\nonumber\\[1 ex]
&=&
\int {\cal D}^8 [ \beta_\kappa ] \, w [ \beta_\kappa ] \, \exp
\left [ \, \pm \, \frac{6}{\gamma \hbar \Lambda} \, \hat{\cal S}_{\rm CS} [ \beta_{\kappa} ] 
\, \right ] \ ,
\label{eq:5.11}
\end{eqnarray}
where 
\begin{equation}
  \label{eq:5.12}
  \hat{\cal S}_{\rm CS} [ \beta_{\kappa} ] := {\cal S}_{\rm CS} [ \omega_{i a} ] -
  {\textstyle \frac{1}{6}} \, I ( \bbox{\hat \Omega} )
\end{equation}
\\
is a locally scale-invariant functional describing the exponent of $| \Psi_{\rm CS} |^2$
according to (\ref{eq:4.16}) and (\ref{eq:4.0.4}). 
The weight function $w [ \beta_\kappa ]$ occuring in (\ref{eq:5.11}) depends 
on the choice of
the new integration variables $\beta_\kappa$. Since the integrand
of (\ref{eq:5.11}) is locally scale-invariant, the integral is independent of the choice 
of $\tilde{\upsilon} (x)$ in
(\ref{eq:5.4}), as announced above, so the gauge-condition $\tilde{\chi} = 0$
can be omitted in the second line of (\ref{eq:5.11}).

As a result, we find that the diffeomorphism- , gauge- and locally scale-invariant 
functional $\hat{\cal S}_{\rm CS} [ \beta_{\kappa} ]$, which is closely related to the
Chern-Simons functional of the Riemannian spin-connection
$\omega_{ia}$, governs the ``probability''-distribution associated with
the Chern-Simons state (\ref{eq:4.16}) in the limit $a_{\cos} \to \infty$. 
Since the functional ${\cal S}_{\rm CS} [ \omega_{i a} ]$ is obviously unbounded from
above and below, we conclude that the norm (\ref{eq:5.11}) cannot be finite, 
even if we fix the remaining gauge-freedoms concerning
the diffeomorphism- and the local $SO(3)$-gauge-transformations.

However, we should keep in mind that the result (\ref{eq:5.11}) has been 
derived for a very special
choice of the gauge-condition $\tilde{\chi}$ according to (\ref{eq:5.4}). Since
different gauge-fixings of the Hamiltonian constraint give rise to \emph{inequivalent}
physical inner products on the Hilbert space of quantum gravity,\footnote{
This is a peculiarity of the Hamiltonian constraint, and in contrast to gauge-fixing
procedures associated with $\tilde{{\cal H}}_i$ or $\tilde{{\cal J}}_a$, for which the
Faddeev-Popov calculus guarantees a \emph{unique} physical inner product 
\cite{FadPop,Wood}.} 
there may still exist other choices of $\tilde{\chi}$, 
for which the Chern-Simons state
$\Psi_{\rm CS} [ \tilde{e}^i \,\!_a ]$ turns out to be normalizable.   


\section{Discussion and Conclusion}
\label{section6}

The main purpose of this paper was to derive and discuss a triad representation of the
Chern-Simons state, which is a well-known exact wavefunctional of quantum gravity
within Ashtekar's theory of general relativity. In particular, we were interested in an explicit 
transformation connecting the real triad representation with the complex Ashtekar 
representation. Therefore, we first investigated this transformation 
on the classical level in section~\ref{section2}. Here we also derived 
new representations for the constraint observables $\tilde{{\cal H}}_0$, $\tilde{{\cal H}}_i$
and $\tilde{{\cal J}}_a$ in terms of a single tensor-density $\tilde{{\cal G}}^i_{\Lambda , a}$
defined in (\ref{eq:2.16}), which is closely related to the curvature ${\cal F}_{i j a}$ 
of the Ashtekar spin-connection ${\cal A}_{i a}$.

Then, in section~\ref{section3}, we performed a canonical quantization of the theory 
in the triad representation.
In the particular factor ordering for the quantum constraint operators $\tilde{{\cal H}}_0$, 
$\tilde{{\cal H}}_i$ and $\tilde{{\cal J}}_a$ suggested by the equations  
(\ref{eq:2.17-})-(\ref{eq:2.17+}) 
we found that the constraint algebra closes formally without any quantum corrections.

On the quantum mechanical level, the transformation from the Ashtekar- to the 
triad representation turned out to be given by a generalized
Fourier transformation (\ref{eq:3.11}) and a subsequent similarity transformation 
(\ref{eq:3.8}). Here it was essential to allow for an arbitrary \emph{complex} integration
manifold $\Gamma$ in the Fourier integral (\ref{eq:3.11}), restricted only by the condition that
partial integrations should be permitted without getting any boundary terms.  

Making use of the transformations (\ref{eq:3.8}) and (\ref{eq:3.11}), we then recovered the
Chern-Simons state of quantum gravity by searching for a wavefunctional which is
annihilated by $\tilde{{\cal G}}^i_{\Lambda , a}$. The 
Chern-Simons state in the triad representation turned out to be given by the 
complex functional integral (\ref{eq:3.17}).
In our approach the Ashtekar variables played only the role of convenient auxiliary
quantities. The \emph{reality conditions} originally introduced by 
Ashtekar in \cite{Ash1} did nowhere enter explicitly, but lie hidden in the choice of the
integration contour $\Gamma$ for the functional integrals in 
(\ref{eq:3.11}) and (\ref{eq:3.17}).  

We did not try to perform the complex functional integral occuring in (\ref{eq:3.17}) 
analytically, but restricted ourselves to semiclassical expansions of the Chern-Simons state, 
which were treated in section~\ref{section4}. Rewriting the state
$\Psi_{\rm CS} [ \tilde{e}^i \,\!_a ]$  in suitable dimensionless 
field-parameters,  the functional integral turned out to be of a Gaussian saddle-point
form in the semiclassical limit $\mu \to \infty$, and the semiclassical  
Chern-Simons state was determined by solutions of the 
saddle-point equations (\ref{eq:4.4}). Here it depended on the choice of the
integration contour $\Gamma$, which particular saddle-points contributed to the
functional integral (\ref{eq:3.17}) via  (\ref{eq:4.2.3}). In order to prove the consistency
of the semiclassical expansions, we argued for the solvability of the saddle-point
equations (\ref{eq:4.4}) in a separate appendix~\ref{appA.1} from a mathematical point
of view, where it turned out, that saddle-point solutions will exist at least under the
restriction $R(x) \not= 2 \Lambda$.   

We were able to find explicit analytical results for the semiclassical Chern-Simons state
in the two asymptotic regimes $\kappa = \Lambda a_{\rm cos}^2 /3 \to \infty$ and
$\kappa \to 0$, which were discussed in sections~\ref{section4.2} and  
\ref{section4.3}, respectively. 

In the limit $\kappa \to \infty$, two different solutions
of the saddle-point equations (\ref{eq:4.4}) could be found, giving rise to the 
linearly independent asymptotic states $\Psi_{\rm CS}$ and  $\Psi^*_{\rm CS}$ given
in (\ref{eq:4.16}). For a suitable choice of the normalization factor ${\cal N}$
according to (\ref{eq:4.0.4}), these asymptotic states turned out to be invariant 
under arbitrary, even topologically non-trivial $SO(3)$-gauge-transformations of the triad.
In the special case of Bianchi-type homogeneous 3-metrics, we obtained the
explicit result (\ref{eq:4.16e}) for the value of the asymptotic Chern-Simons 
state (\ref{eq:4.16}), which, by a further restriction to Bianchi-type~IX metrics, 
coincided with the corresponding result known from 
discussions of the homogeneous Bianchi-type~IX model. 

The asymptotic Chern-Simons state (\ref{eq:4.16}) in the limit $\kappa \to \infty$
gives rise to a well-defined 
semiclassical time-evolution, which we discussed in section~\ref{section4.2.2}. There
it turned out, that for large scale parameters $a_{\rm cos}$ the semiclassical 4-geometries 
associated with the Chern-Simons state are given by inhomogeneously generalized
deSitter space-times.

In the limit $\kappa \to 0$, the semiclassical saddle-point contributions to the
Chern-Simons state can be characterized by divergence-free triads $\vec{d}_a$ of the
Riemannian 3-manifold $( {\cal M}_3 , \bbox{h} )$ via (\ref{eq:4.31}). Thus we had to
answer the non-trivial question, whether divergence-free triads to a given 3-metric
will in general \emph{exist}, what was done in appendix~\ref{appA.2}. 

In restriction
to homogeneous manifolds of Bianchi-type~A, \emph{one} divergence-free triad was
explicitly known, giving rise to the result (\ref{eq:4.32.1}). In particular, we were
able to recover the ``wormhole-state'' (\ref{eq:4.32.2}), which is a well-known 
vacuum state within the homogeneous Bianchi~IX model. For 
Bianchi-type~IX manifolds, four further divergence-free triads
$\vec{d}^{\, ( \alpha )}_a , \alpha \in \{ 1,2,3,4 \},$ were constructed in appendix~\ref{appB}.
They gave rise to four additional saddle-point contributions 
$\Psi_{\rm vac}^{ ( \alpha )} \, , \alpha \in \{ 1,2,3,4 \},$ to the vacuum Chern-Simons state,
which, however, were restricted to
occur simultaneously. We concluded that, together with the ``wormhole-state'', 
only \emph{two} linearly independent values of the vacuum Chern-Simons state are
realized for Bianchi-type~IX manifolds. 

Since these two values should continue to
exist under sufficiently small, inhomogeneous perturbations of the 3-metric, and since
also in the limit $\kappa \to \infty$ exactly \emph{two} different values
of the semiclassical Chern-Simons state were found, one may assume that the one
Chern-Simons state in the Ashtekar representation corresponds to \emph{two} linearly
independent states in the triad representation.

Within the narrow class of Bianchi-type~IX
metrics, the semiclassical 4-geometries associated with the vacuum Chern-Simons state
(\ref{eq:4.31}) are satisfying physically interesting boundary conditions,
namely either the ``no-boundary'' condition proposed by Hartle and Hawking 
\cite{Haw1,Haw2,Haw3}, or the condition of asymptotical flatness at large scale
parameters $a_{\rm cos}$. However, this does \emph{not} remain true for general
3-metrics, as we have shown by exhibiting a counter-example in appendix~\ref{appC}. 
We conclude that, in general, the Chern-Simons state will not satisfy the ``no-boundary''
condition or the condition of asymptotical flatness. Nevertheless, as we have 
remarked in section~\ref{section5}, the asymptotic state (\ref{eq:4.16}) in the
limit $\kappa \to \infty$ reminds one of the \emph{Vilenkin} proposal for the 
wavefunction of the Universe \cite{Vil1,Vil3}.

In section~\ref{section5}, we investigated the normalizability of the Chern-Simons
state (\ref{eq:3.17}) in the triad representation. We defined a kinematical inner product
on the Hilbert space of quantum gravity, and by performing a special gauge-fixing for the
time-gauge we arrived at the physical inner product (\ref{eq:5.2}). Unfortunately,
the Chern-Simons state turned out to be \emph{non-normalizable} with respect to
this particular inner product.
However, as we have pointed out, there may still exist other gauge-fixing procedures
(e.g. the one suggested by Smolin and Soo in \cite{Smo}),
which render the Chern-Simons state to be normalizable.


\acknowledgements

Support of this work by the Deutsche Forschungsgemeinschaft through the 
Sonderforschungsbereich ``Unordnung und gro{\ss}e Fluktuationen'' is gratefully
acknowledged. We further wish to thank Prof.~Abresch from the Ruhr-Universit\"at
Bochum for many fruitful discussions and important ideas concerning the mathematical
problems discussed in appendix~\ref{appA}.


\begin{appendix}


\section{On the solvability of the saddle-point equations}
\label{appA}

The solvability of the semiclassical saddle-point equations (\ref{eq:4.4}) is essential
in order to justify the consistency of the asymptotical expansions of the Chern-Simons state 
discussed in section~\ref{section4}. Therefore, it is worth to study the solvability
properties of the nonlinear, partial differential equations (\ref{eq:4.4}) from a 
mathematical point of view, what will be done in section~\ref{appA.1}. Applying
the results of section~\ref{appA.1} to the special case of a vanishing cosmological constant
$\Lambda$, we will then, in section~\ref{appA.2}, be able to prove the
existence of divergence-free triads of Riemannian 3-manifolds, 
which determine the semiclassical vacuum state (\ref{eq:4.31}).


\subsection{The general case $\Lambda \not= 0$}
\label{appA.1}

If we want to discuss the solvability of the saddle-point equations (\ref{eq:4.4}) within the
theory of partial differential equations (cf.~\cite{Rau}), 
it is \emph{not} advisable to study this problem 
in the particular form (\ref{eq:4.4}), since the spatial derivative operator, which is given
by the curl of the gauge-field ${\cal A}_{i a}$, is known to be \emph{non-elliptic}.
However, we will show that it is possible to consider a set of second order partial
differential equations instead, which will turn out to be elliptic in leading derivative order, 
thus allowing for solvability statements concerning the solutions ${\cal A}_{i a}$.

Let us first introduce new variables
\begin{equation}
  \label{eq:a.1}
  {\cal K}_{i j} := \left ( \omega_{i a} - {\cal A}_{i a} \right ) e_{j a} \equiv \mp i \, K_{j i} 
\end{equation}
instead of the gauge-fields ${\cal A}_{i a}$, where $e_{i a}$ denotes a fixed triad
for which we want to solve the set of equations (\ref{eq:4.4}). 
Up to a Wick-rotation, the tensor
${\cal K}_{i j}$ plays the role of the semiclassical extrinsic curvature tensor $K_{i j}$
(cf.~eqs.~(\ref{eq:2.2}), (\ref{eq:2.6}) and (\ref{eq:2.11})).
If we rewrite the saddle-point equations (\ref{eq:4.4}) in terms of the new
variables ${\cal K}_{i j}$, they become
\begin{equation}
  \label{eq:a.2}
  {\cal G}^i_{\Lambda , j} := \frac{1}{\sqrt{h}} \  \tilde{{\cal G}}^i_{\Lambda , a} \, e_{j a} =
  G^i_{\Lambda , j} + {^* {{\cal K}^i}_j} - \frac{1}{\sqrt{h}} \, \tilde{\varepsilon}^{i k \ell} \,
  \nabla_{\! k} \, {\cal K}_{\ell j} \stackrel{!}{=} 0 \ , 
\end{equation}
where
\begin{equation}
  \label{eq:a.3}
  {^* {{\cal K}^i}_j} := {\textstyle \frac{1}{2}} \, \tilde{\varepsilon}^{i k \ell} \,
  \, \mbox{\raisebox{-1.9 ex}{$\tilde{}$} \hspace{-1.2 ex}} \varepsilon_{j m n} \,
  {{\cal K}_k}^m \, {{\cal K}_{\ell}}^{\, n} 
\end{equation}
are the cofactors of the matrix-elements ${{\cal K}_i}^{\, j}$, and $G^i_{\Lambda , j}$ is the
usual, 3-dimensional Einstein-tensor with a cosmological term. In analogy to 
(\ref{eq:2.17+}), the set of equations (\ref{eq:a.2}) implies the three Gau{\ss}-constraints
\begin{eqnarray}
  \tilde{{\cal J}}_a & = &
  \pm \, \frac{ 6 i}{\gamma \Lambda} \left [
  \nabla_{\! j} \, {\cal G}^j_{\Lambda , i} - \sqrt{h} \  
  \, \mbox{\raisebox{-1.9 ex}{$\tilde{}$} \hspace{-1.2 ex}} \varepsilon_{i j k} \,
  \, {{\cal K}_{\ell}}^{\, j} \, {\cal G}^{\ell k}_{\Lambda} \right ]
  \tilde{e}^i \,\!_a  \nonumber \\[1 ex]
  & \equiv & \pm \, \frac{ 2 i}{\gamma} \, e_{i a}
  \, \tilde{\varepsilon}^{i j k} \, {\cal K}_{j k}
  \stackrel{!}{=} 0 \ ,
\label{eq:a.4}
\end{eqnarray}
which require the tensor ${\cal K}_{i j}$ to be \emph{symmetric} in $i$ and $j$. Therefore,
if we take ${\cal K}_{i j}$ \emph{to be} symmetric in the following, the Gau{\ss}-constraints
(\ref{eq:a.4}) are satisfied identically, and the first line of (\ref{eq:a.4}) takes the form of
three generalized Bianchi-identities. We thus conclude that the set of equations
(\ref{eq:a.2}) constitutes only \emph{six} independent equations for the \emph{six}
fields ${\cal K}_{i j} = {\cal K}_{j i}$ we are searching for. 

Beside the Gau{\ss}-consraints (\ref{eq:a.4}), four further equations are implied by 
(\ref{eq:a.2}) via (\ref{eq:2.17-}) and (\ref{eq:2.17}), namely the Hamiltonian constraint
\begin{equation}
  \label{eq:a.5}
  \tilde{{\cal H}}^{\rm ADM}_0 = \frac{2 \sqrt{h}}{\gamma} \ {\cal G}^i_{\Lambda , i} 
  \equiv \frac{\sqrt{h}}{\gamma} \Bigl ( {\cal K}^2 - {{\cal K}^i}_j \, {{\cal K}^j}_i +
  2 \Lambda - R \Bigr ) \stackrel{!}{=} 0 \ ,
\end{equation}
and the three diffeomorphism-constraints
\begin{equation}
  \label{eq:a.6}
  \tilde{{\cal H}}_i = \mp \, \frac{2 i h}{\gamma} \ 
  \, \mbox{\raisebox{-1.9 ex}{$\tilde{}$} \hspace{-1.2 ex}} \varepsilon_{i j k} \,
  {\cal G}^{j k}_{\Lambda}
  \equiv \pm \frac{2 i \sqrt{h}}{\gamma} \, \Bigl ( \nabla_{\! j} \, {{\cal K}^j}_i - \nabla_{\! i} \,
  {\cal K} \Bigr ) \stackrel{!}{=} 0 \ ,
\end{equation}
respectively. Here ${\cal K}$ in (\ref{eq:a.5}) and (\ref{eq:a.6}) 
denotes the trace of $( {{\cal K}^i}_j )$. Remarkably, the Hamiltonian constraint (\ref{eq:a.5})
is a purely algebraical equation for ${\cal K}_{i j}$, which will be solved explicitly later on,
while the diffeomorphism-constraints (\ref{eq:a.6}) are \emph{linear} equations and
contain information about the \emph{divergence} of the fields ${\cal K}_{i j}$. 

Moreover, since the equations (\ref{eq:a.2}) contain the curl of the fields ${\cal K}_{i j}$,
eqs.~(\ref{eq:a.2}) and (\ref{eq:a.6}) together may be used to construct
a second order derivative operator similar to the Laplace-Beltrami-operator of 
${\cal K}_{i j}$.  Let us therefore consider the following second order differential equations
\begin{equation}
  \label{eq:a.7}
  \Delta_{i j} := \sqrt{h} \left [
  \, \mbox{\raisebox{-1.9 ex}{$\tilde{}$} \hspace{-1.2 ex}} \varepsilon_{j m n} \,
  \nabla_{\! i} \, {\cal G}^{m n}_{\Lambda} -
  \, \mbox{\raisebox{-1.9 ex}{$\tilde{}$} \hspace{-1.2 ex}} \varepsilon_{i m n} \,    
  h^{m k} \, \nabla_{\! k} \, {\cal G}^n_{\Lambda , j} + {\textstyle \frac{1}{2}} \,
  \, \mbox{\raisebox{-1.9 ex}{$\tilde{}$} \hspace{-1.2 ex}} \varepsilon_{i j k} \,
  \nabla_{\! n} \, {\cal G}^{n k}_{\Lambda} \right ] \stackrel{!}{=} 0 \ ,
\end{equation}
which must be satisfied for solutions ${\cal K}_{i j}$ of (\ref{eq:a.2}).
The first term in (\ref{eq:a.7}) can be simplified with help of (\ref{eq:a.6}), and gives
in the leading derivative order the gradient of the divergence of ${\cal K}_{i j}$ and,
in addition, the Hessian of ${\cal K}$. Making use of eqs.~(\ref{eq:a.2}), 
the second term in (\ref{eq:a.7}) 
contributes the curl of the curl of ${\cal K}_{i j}$, i.e. taking the first two terms
in (\ref{eq:a.7}) together, we arrive at 
\begin{equation}
  \label{eq:a.8}
  \Delta_{i j} = \nabla_{\! i} \nabla_{\! j} \, {\cal K} - \Delta \, {\cal K}_{i j} +
  {\cal O} ( \nabla_{\! i} \, {\cal K}_{j k} )
\end{equation}
in leading derivative order. By virtue of eqs.~(\ref{eq:a.4}), the
third term in (\ref{eq:a.7}) contains only first order derivatives of ${\cal K}_{i j}$. It has
been added to obtain simple expressions for the trace and the antisymmetric part of
$\Delta_{i j}$, which are given by
\begin{equation}
  \label{eq:a.9}
  \qquad h^{i j} \, \Delta_{i j} \equiv 0 \qquad , \qquad
  \tilde{\varepsilon}^{i j k} \, \Delta_{j k} \equiv \frac{\gamma}{2} \, h^{i j} \, \nabla_{\! j} \,
  \tilde{{\cal H}}^{\rm ADM}_0 \ .
\end{equation}
Instead of solving the nine equations (\ref{eq:a.7}), we may therefore consider the
six equations
\begin{equation}
  \label{eq:a.10}
  \Delta_{( i j )} := {\textstyle \frac{1}{2}} \left ( \Delta_{i j} + \Delta_{j i} \right ) \stackrel{!}{=} 0
  \qquad , \qquad \tilde{{\cal H}}^{\rm ADM}_0 \stackrel{!}{=} 0 
\end{equation}
to determine the six fields ${\cal K}_{i j}$.

In a next step, we will now solve the Hamiltonian constraint (\ref{eq:a.5}) explicitly. At
any space-point $x \in {\cal M}_3$, eq.~(\ref{eq:a.5}) describes a five dimensional
hyperboloid in the six dimensional space spanned by ${\cal K}_{i j}$, as long as 
\begin{equation}
  \label{eq:a.11}
  \forall x \in {\cal M}_3 : \ R(x) \not= 2  \Lambda \ , 
\end{equation}
which will be assumed in the following. This five dimensional hyperboloid may be
parameterized with help of a stereographic projection, hence the general solution of
the Hamiltonian constraint can be written in the form
\begin{equation}
  \label{eq:a.12}
  {{\cal K}^i}_j = \frac{\sqrt{R - 2 \Lambda}}{1 - \mbox{Tr}  \bbox{{\cal Q}}^2} 
  \left [ \frac{1 + \mbox{Tr}  \bbox{{\cal Q}}^2}{\sqrt{6}} \, \delta^i_j + 2 \, {{\cal Q}^i}_j
  \right ] \qquad , \qquad \mbox{Tr}  \bbox{{\cal Q}}^2 \not= 1 \ , 
\end{equation}
\\
where $\bbox{{\cal Q}}$ is a symmetric, \emph{traceless} matrix. Matrices $\bbox{{\cal Q}}$
with $\mbox{Tr} \bbox{{\cal Q}}^2 = 1$ correspond to coordinate singularities of the
stereographic projection, and thus have to be excluded in (\ref{eq:a.12}). Inserting the 
general solution (\ref{eq:a.12}) of $\tilde{{\cal H}}^{\rm ADM}_0 = 0$ into the first of 
eqs.~(\ref{eq:a.10}), we arrive at \emph{five} equations for the \emph{five}
fields ${{\cal Q}^i}_j$, which remain to be determined. 

We now want to argue that the effective set of partial differential equations obtained
this way is soluble with respect to ${{\cal Q}^i}_j$. Let us therefore consider a background
solution ${{\bar{{\cal Q}}}^i}_{\ j}$ of these equations, which we assume to be known for 
sufficiently simple parameter fields $\tilde{e}^i \,\!_a$ and $\Lambda$.\footnote{
Explicit solutions ${\cal A}_{i a}$ of the saddle-point eqs.~(\ref{eq:4.4}), 
which correspond to 
the fields ${{\cal Q}^i}_j$ via (\ref{eq:a.1}) and (\ref{eq:a.12}), are in fact known for various
\emph{homogeneous} 3-manifolds, such as Bianchi-type~IX manifolds, cf.~\cite{Pat1}.}
Under infinitesimal perturbations of the parameter fields $\tilde{e}^i \,\!_a$ and $\Lambda$,
the new solution ${{\cal Q}^i}_j$ will differ from the background solution 
${{\bar{{\cal Q}}}^i}_{\ j}$ by an infinitesimal
amount 
\begin{equation}
  \label{eq:a.13}
  {{\cal Q}^i}_j = {{\bar{\cal Q}}^i}_{\ j} + \epsilon \cdot {{{\cal Q}'}^i}_j + {\cal O} ( \epsilon^2 ) \ ,
\end{equation}
and in the following it will be sufficient to show that the fields ${{{\cal Q}'}^i}_j$ 
exist to any given background solution ${{\bar{\cal Q}}^i}_{\ j}$. Inserting the perturbation
ansatz (\ref{eq:a.13}) into $\Delta_{(i j)} = 0$, we arrive at five \emph{linear} partial
differential equations $\Delta' \,\!\!_{(i j)} =0$ in ${\cal O} ( \epsilon )$ determining the fields
${{{\cal Q}'}^i}_j$. To show that these equations are soluble with respect to
${{{\cal Q}'}^i}_j$, we will restrict ourselves to a discussion of the \emph{symbol} of
$\Delta' \,\!\!_{(i j)} =0$, which we will show to be \emph{elliptic} (cf.~\cite{Rau}). The symbol 
$\sigma ( \bbox{k} )$ of a linear
differential operator is obtained by computing the action on a Fourier mode
\begin{equation}
  \label{eq:a.14}
  {{{\cal Q}'}^i}_j (x) = {{\hat{{\cal Q}}}^i}_{\; j} (\bbox{k} ) \cdot e^{i \, k_{\ell} \, x^{\ell}} \ ,
\end{equation}
in leading order of the wavevector $\bbox{k}$. For the operator $\Delta' \,\!\!_{(i j)}$ under
study, we obtain
\begin{eqnarray}
 \sigma_{i j} \left ( \Delta'_{(m n)} ; \bbox{k} \right ) & = &
 -2 \, \frac{{\sqrt{R - 2 \Lambda}}}{ \left ( 1 - \mbox{Tr} {{\bbox{\bar{{\cal Q}}}}}^2 \right )^2}
 \, \Bigg [ \sqrt{6} \, k_i \, k_j \, {\bar{{\cal Q}}}^{m n} \, {\hat{{\cal Q}}}_{m n} - | \bbox{k} |^2 
  \Bigg ( \left ( 1- \mbox{Tr} {{\bbox{\bar{{\cal Q}}}}}^2 \right ) {\hat{{\cal Q}}}_{i j}
  \nonumber \\
  && \qquad \qquad \qquad \qquad 
  + \sqrt{\frac{2}{3}} \  {\bar{{\cal Q}}}^{m n} \left ( h_{i j} + \sqrt{6} \, 
  {\bar{{\cal Q}}}_{i j} \right ) {\hat{{\cal Q}}}_{m n} \Bigg ) \Bigg ] \ .
 \label{eq:a.15}
\end{eqnarray}
\\
The symbol $\sigma ( \bbox{k} )$ is called elliptic, if it has a \emph{trivial} kernel for
$\bbox{k} \not= \bbox{0}$. Then the linear differential operator is invertable in the
leading derivative order, and solutions of the linear differential equations will exist.
To prove the ellipticity of the symbol (\ref{eq:a.15}), it remains to be shown that
the linear equations
\begin{eqnarray}
  \label{eq:a.16}
  \sqrt{6} \, q \, n_i \, n_j = \sqrt{\frac{2}{3}} \, q \left ( h_{i j} + 
  \sqrt{6} \, {\bar{{\cal Q}}}_{i j} \right )
  + \left ( 1 - \mbox{Tr} {\bbox{\bar{{\cal Q}}}}^2 \right ) \, {\hat{{\cal Q}}}_{i j} 
\end{eqnarray}
have only the trivial solution ${\hat{{\cal Q}}}_{i j} = 0$ for $\bbox{n} \not= \bbox{0}$, where
we have introduced the abbreviations
\begin{eqnarray}
  \label{eq:a.17}
  q := {\bar{{\cal Q}}}^{i j} \, {\hat{{\cal Q}}}_{i j} \qquad , \qquad
  \bbox{n} := \frac{\bbox{k}}{| \bbox{k} |} \ \ \Rightarrow \ \ | \bbox{n} | = 1 \ .
\end{eqnarray}
Contracting eqs.~(\ref{eq:a.16}) with ${\bar{{\cal Q}}}^{i j}$, we obtain the 
necessary implication
\begin{equation}
  \label{eq:a.18}
  q \left ( 1 + \mbox{Tr} {\bbox{\bar{{\cal Q}}}}^2 - \sqrt{6} \, {\bar{{\cal Q}}}^{i j} \, n_i \, n_j
  \right ) \stackrel{!}{=} 0 \ ,
\end{equation}
i.e. if we can show that the bracket in (\ref{eq:a.18}) is different from zero, 
eq.~(\ref{eq:a.18})
implies $q=0$, and therefore ${\hat{{\cal Q}}}_{i j} = 0$ via (\ref{eq:a.16}), so the ellipticity
of $\sigma ( \bbox{k} )$ according to (\ref{eq:a.15}) would have been proven.

It now follows from a simple estimate for symmetric matrices $\bbox{\bar{{\cal Q}}}$ that
the vanishing of the bracket in (\ref{eq:a.18}) implies\footnote{
Here and in the following, we have to restrict ourselves to \emph{real-valued} matrices
$\bbox{\bar{{\cal Q}}}$, which correspond to real or complex solutions ${\cal A}_{i a}$ of
the saddle-point equations (\ref{eq:4.4}) via (\ref{eq:a.12}) and (\ref{eq:a.1}) in the
two different cases $R > 2 \Lambda$ or $R < 2 \Lambda$, respectively.}  
\begin{equation}
  \label{eq:a.19}
  1 + \sum_{i=1}^3 \, {\bar{{\cal Q}}}^2_i \leq \sqrt{6} \, \max_{i=1}^3 \, \{ {\bar{{\cal Q}}}_i \} \,
\end{equation}
where the ${\bar{{\cal Q}}}_i$ denote the three eigenvalues of the matrix 
$\bbox{\bar{{\cal Q}}}$.
Since $\bbox{\bar{{\cal Q}}}$ is traceless, these three eigenvalues may be parameterized
by
\begin{equation}
  \label{eq:a.20}
  {\bar{{\cal Q}}}_j = \sqrt{\frac{2}{3}} \, \varrho \, \cos \left ( \theta + \frac{2 \pi j}{3} \right ) \ \ 
  , \ \ j \in \{ 1,2,3 \} \qquad \mbox{with} \qquad \varrho \geq 0 \ \ , \ \ 0 \leq \theta < 2 \pi \ .
\end{equation}
Then the relation (\ref{eq:a.19}) takes the form
\begin{equation}
  \label{eq:a.21}
  1 + \varrho^2 \leq 2 \, \varrho \qquad \Leftrightarrow \qquad ( 1 - \varrho )^2 \leq 0 \ ,  
\end{equation}
and is obviously only satisfied for $\varrho = 1$. Moreover, because of the identity
$\mbox{Tr} {\bbox{\bar{{\cal Q}}}}^2 = \varrho^2$, the particular value $\varrho = 1$
corresponds to the coordinate singularity of the stereographic projection used in
(\ref{eq:a.12}), and is hence not permitted by construction. Thus 
the relation (\ref{eq:a.19}) has been brought to a contradiction, and we conclude that
the bracket in (\ref{eq:a.18}) cannot vanish, what finishes our proof of the ellipticity
of the symbol $\sigma ( \bbox{k} )$ given in (\ref{eq:a.15}).

Summarizing our results, we have shown that the set of linear partial differential
equations $\Delta' \,\!\!_{(i j)} = 0$, which determines the fields ${{{{\cal Q}}'}^i}_j$, is elliptic,
and therefore soluble in leading derivative order. It follows, that the solutions 
${{{\cal Q}}^i}_j$ of the nonlinear set of equations $\Delta_{(i j)} = 0$ continue to exist
under infinitesimal perturbations of the parameter fields $\tilde{e}^i \,\!_a$ and $\Lambda$.
Therefore, solutions ${\cal K}_{i j}$ of (\ref{eq:a.7}), and also solutions ${\cal A}_{i a}$ of
the saddle-point equations (\ref{eq:4.4}) can be obtained via (\ref{eq:a.12}) and
(\ref{eq:a.1}) for a wide range of parameter fields $\tilde{e}^i \,\!_a$ and
$\Lambda$, as long as the only restriction $R \not= 2 \Lambda$ met in (\ref{eq:a.11})
is satisfied.


\subsection{Divergence-free triads in the limit $\Lambda \to 0$}
\label{appA.2}

In this section we want to discuss how suitable \emph{flat} gauge-fields ${\cal A}_{i a}$
may be used to construct divergence-free triads $\vec{d}_a$ of a given
Riemannian 3-manifold $({\cal M}_3 ,  \bbox{h})$.
Such a flat gauge-field on ${\cal M}_3$ can be obtained by 
pursuing any \emph{fixed} solution 
${\cal A}_{i a} [ \tilde{e}^i \,\!_a , \Lambda ]$ of the saddle-point equations
(\ref{eq:4.4}) in the limit $\Lambda \to 0$. Using the arguments of section~\ref{appA.1},
this will be possible for 3-manifolds with $R(x) \not= 0$.
By virtue of (\ref{eq:2.17+}), the corresponding gauge-field ${\cal A}_{i a}$ 
will not only be flat, but it
will in addition satisfy the three Gau{\ss}-constraints  
\begin{equation}
  \label{eq:a.22}
  {\cal D}_i \, \tilde{e}^i \,\!_a \equiv \partial_i \tilde{e}^i \,\!_a + 
  \varepsilon_{a b c} \, {\cal A}_{i b} \, \tilde{e}^i \,\!_c = 0 \ ,
\end{equation}
where $\tilde{e}^i \,\!_a$ is a fixed but arbitrary triad of the 3-metric $\bbox{h}$. 

Let us now consider the \emph{parallel transport} associated with the gauge-field 
${\cal A}_{i a}$: Given a vector $\vec{v} (0) = v_{a , 0} \, \vec{e}_a$ at a point $P_0$ 
of ${\cal M}_3$, and a curve ${\cal C}: \, x^i = f^i (u) \, , 0 \leq u \leq 1$, 
connecting $P_0$ with a 
second point $P_1$, we define a vector-field $\vec{v} (u)$ along ${\cal C}$ by solving
the equations of parallel transport, 
\begin{equation}
  \label{eq:a.23}
  \frac{{\cal D} v_a}{{\cal D} u} := \frac{\partial v_a}{\partial u} +
  \varepsilon_{a b c} \, \frac{\partial f^i}{\partial u} \, {\cal A}_{i b} \, v_c \stackrel{!}{=} 0 
  \qquad , \qquad v_a (0) \stackrel{!}{=} v_{a , 0} \ .
\end{equation}
\\
Since the gauge-field ${\cal A}_{i a}$ is flat, the resulting vector $\vec{v} (1)$ at the
Point $P_1$ does \emph{not} depend on the particular choice of ${\cal C}$
(cf.~\cite{RySch}), i.e. if we
restrict ourselves to the case of \emph{simply connected} manifolds ${\cal M}_3$
in the following, the parallel
transport of $\vec{v} (0)$ along arbitrary curves ${\cal C} \subset {\cal M}_3$ will
define a well-defined \emph{vector-field} $\vec{v} (x)$ on ${\cal M}_3$. By construction,
this vector-field $\vec{v} (x)$ turns out to be covariantly constant with respect to
${\cal A}_{i a}$,
\begin{equation}
  \label{eq:a.24}
  {\cal D}_i \, v_a \equiv \partial_i v_a + \varepsilon_{a b c} \, {\cal A}_{i b} \, v_c \equiv 0 \ ,
\end{equation}
and, as a consequence of eq.~(\ref{eq:a.22}), the vector-field $\vec{v} (x)$ is in addition 
\emph{divergence-free},
\begin{equation}
  \label{eq:a.25}
  \vec{\nabla} \cdot \vec{v} \equiv \frac{1}{\sqrt{h}} \, {\cal D}_i \! \left (
  v_a \, \tilde{e}^i \,\!_a \right ) = \frac{1}{\sqrt{h}} \, \Big (
  \underbrace{{\cal D}_i \, v_a}_0 \, \tilde{e}^i \,\!_a +
  v_a \, \underbrace{{\cal D}_i \, \tilde{e}^i \,\!_a}_0 \Big ) = 0 \ . 
\end{equation}
Moreover, it follows from eq.~(\ref{eq:a.24})
that the parallel transport according to (\ref{eq:a.23}) conserves the 
scalar product of two vectors $\vec{v}$ and $\vec{w}$:
\begin{equation}
  \label{eq:a.26}
  \partial_i \left ( \vec{v} \cdot \vec{w} \right )
  \equiv {\cal D}_i \! \left ( v_a \, w_a \right )
  = \underbrace{{\cal D}_i \, v_a}_0 \, w_a + v_a \, \underbrace{{\cal D}_i \, w_a}_0 = 0 \ .
\end{equation}

From eqs.~(\ref{eq:a.25}) and (\ref{eq:a.26}) it is then obvious that 
a divergence-free \emph{triad} $\vec{d}_a (x)$ of the
Riemannian 3-manifold $({\cal M}_3 , \bbox{h} )$ can be constructed
by choosing three orthonormal vectors $\vec{d}_a$ at a point $P_0$, and 
parallel-propagating
these vectors along arbitrary curves ${\cal C} \subset {\cal M}_3$. 
Since the only freedom in this construction arises from the 
choice of $\vec{d}_a$ at a single point $P_0$, this divergence-free triad $\vec{d}_a (x)$ 
associated with the flat gauge-field ${\cal A}_{i a}$ turns out to be unique up 
to \emph{global} rotations.


\section{The vacuum state on Bianchi-type~IX homogeneous manifolds}
\label{appB}

In this appendix we want to discuss the semiclassical vacuum state (\ref{eq:4.31}) in the
special case of Bianchi-type~IX homogeneous 3-manifolds. While one saddle-point
contribution, the so-called ``wormhole-state'', is given by the result (\ref{eq:4.32.2}), four
further semiclassical vacuum states are known within the framework of the
homogeneous Bianchi~IX model
\cite{Pat1,Pat3}. In the inhomogeneous approach of the present paper, these 
additional states should correspond to topologically nontrivial divergence-free
triads of Bianchi-type~IX manifolds via (\ref{eq:4.31}). Such special 
triads can indeed be constructed from the divergence-free triads of the unit 3-sphere,
which will be discussed first in section~\ref{appB.1}. The divergence-free triads
of Bianchi-type~IX manifolds and the corresponding saddle-point contributions to the
vacuum Chern-Simons state will then be given in section~\ref{appB.2}.


\subsection{Divergence-free triads of the unit 3-sphere}
\label{appB.1}

The 3-sphere is a maximally symmetric 3-manifold with six killing-vectors 
$\vec{\xi}^{\, \pm}_a$, representing the commutator algebra
\begin{equation}
  \label{eq:b.1}
  \left [ \vec{\xi}^{\, \pm}_a \, , \, \vec{\xi}^{\, \pm}_b \right ] = \pm 2 \, [ a b c ] \,
  \vec{\xi}^{\, \pm}_c \qquad , \qquad 
  \left [ \vec{\xi}^{\, +}_a \, , \, \vec{\xi}^{\, -}_b \right ] = \vec{0}
\end{equation}
of the symmetry group $SO(4) \cong SO(3) \times SO(3)$. 
From the second of these commutation 
relations it follows that the three vector-fields $\vec{\xi}^{\, -}_a$ are the left-invariant
vector-fields to the killing-vectors $\vec{\xi}^{\, +}_a$, and vice versa, i.e. the metric
tensor of the unit 3-sphere can be expanded in \emph{both} of the two sets
$\vec{\xi}^{\, \pm}_a$ with \emph{spatially constant} coefficients. In particular, if we 
choose the normalization of $\vec{\xi}^{\, \pm}_a$ as in the first of eqs.~(\ref{eq:b.1}),
the invariant vector fields $\vec{\xi}^{\, \pm}_a$ form automatically two different sets of 
\emph{invariant triads} $\vec{\imath}^{\; \pm} \!\!\!\!_a := \vec{\xi}^{\, \pm}_a$ 
to the metric $\bbox{h}$ of the unit 3-sphere:   
\begin{equation}
  \label{eq:b.2}
  \vec{\imath}^{\; +} \!\!\!\!_a \otimes \vec{\imath}^{\; +} \!\!\!\!_a = \bbox{h}
 = \vec{\imath}^{\; -} \!\!\!\!_a \otimes \vec{\imath}^{\; -} \!\!\!\!_a \ . 
\end{equation}
According to (\ref{eq:b.1}) and (\ref{eq:4.16a}), both invariant triads 
${\vec{\imath}^{\; \pm} \!\!\!\!_a}$ have a
\emph{symmetric} structure matrix $\bbox{m}$, and
are thus \emph{divergence-free} by virtue of eq.~(\ref{eq:4.32}). Since they are triads
to the same metric $\bbox{h}$, they must be connected by a gauge-transformation
$\bbox{E} \in O(3)$:
\begin{equation}
  \label{eq:b.3}
  \vec{\imath}^{\; +} \!\!\!\!_a = E_{a b} \, 
  \vec{\imath}^{\; -} \!\!\!\!_b \ .
\end{equation}
The matrix $\bbox{E}$ has a spatially nontrivial dependence, and may of course be calculated
explicitly in any given coordinate system on $S^3$.\footnote{For example, if we
employ the Euler angles $\psi , \vartheta, \varphi$ as coordinates on the unit 3-sphere,
the matrix $\bbox{E}$ turns out to be precisely the well-known Euler-matrix 
$\bbox{E} ( \psi , \vartheta , \varphi )$ (for a definition of the Euler-matrix, 
see e.g.~\cite{Gold}).}
However, in the following the explicit form of the rotation
matrix $\bbox{E}$ will not be needed.


\subsection{Divergence-free triads of Bianchi-type~IX homogeneous manifolds}
\label{appB.2}

Anisotropic manifolds of Bianchi-type~IX can be described by choosing an
invariant triad of the unit 3-sphere, for example $\vec{\imath}^{\; +} \!\!\!\!_a$, and 
rescaling this triad with three scale parameters ${a}_{b} > 0$:
\begin{equation}
  \label{eq:b.4}
  \vec{\imath}_a := D_{a b} \, \vec{\imath}^{\; +} \!\!\!\!_b \qquad \mbox{with} \qquad
  \bbox{D}^{-1} := \mbox{diag} \left ( a_1 , a_2 , a_3 \right ) \ .
\end{equation}
Then $\vec{\imath}_a$ is the invariant triad of a 
Bianchi-type~IX manifold, and the metric tensor
is given by $\bbox{h} = \vec{\imath}_a \otimes \vec{\imath}_a$.  
In the general, anisotropic case, only three of the six vector-fields $\vec{\xi}^{\, \pm}_a$
discussed in section~\ref{appB.1}
remain as killing-vectors of the 3-metric $\bbox{h}$, namely the fields $\vec{\xi}^{\, -}_a$.
We will assume that the invariant triad $\vec{\imath}_a$ given in (\ref{eq:b.4}) 
is positive-oriented.
As pointed out in section~\ref{section4.3.1}, this triad 
$\vec{d}^{\, (0)}_a := \vec{\imath}_a$ is automatically divergence-free, and gives rise
to the ``wormhole'' saddle-point contribution (\ref{eq:4.32.2}) to the semiclassical vacuum
state. 

To find further, topologically nontrivial divergence-free triads $\vec{d}_a$ 
of Bianchi-type~IX metrics, let us try an ansatz of the form
\begin{equation}
  \label{eq:b.5}
  \vec{d}_a = E_{b a} \, O_{b c} \, \vec{\imath}_c \ ,
\end{equation}
where $\bbox{O} = ( O_{a b} ) \in SO(3)$ is assumed to be spatially constant.\footnote{
At least in the isotropic case $a_1 = a_2 = a_3$, this ansatz gives the second 
divergence-free triad $\vec{\imath}^{\, -} \!\!\!\!_a$ of the 3-sphere by virtue of
(\ref{eq:b.3}), if we simply choose $\bbox{O} = \bbox{1}$.}
If we require the triad $\vec{d}_a$ according to (\ref{eq:b.5}), (\ref{eq:b.4}) to be 
divergence-free, we arrive at three equations for the matrix~$\bbox{O}$,
\begin{equation}
  \label{eq:b.6}
  \vec{\nabla} \cdot \vec{d}_a = O_{b c} \, D_{c d} \, [ \,  
  \vec{\imath}  \raisebox{1.2 ex}{\mbox{\scriptsize{$\; +$}}} \!\!\!\!_d \, , 
  \, E_{b a} \, ]
  \stackrel{!}{=} 0 \ .
\end{equation}
The spatial derivatives of the
matrix $\bbox{E}$ with respect to the vector-fields $ \vec{\imath}^{\; +} \!\!\!\!_a$ can
be calculated by inserting eqs.~(\ref{eq:b.3}) into eqs.~(\ref{eq:b.1}), and are given by
\begin{equation}
  \label{eq:b.7}
  [ \,  \vec{\imath}^{\; +} \!\!\!\!_a \, , \, E_{b c} \, ] = 2 \, \varepsilon_{a b d} \, E_{d c} \ .
\end{equation}
Therefore, the requirements (\ref{eq:b.6}) can be simplified to the form
\begin{equation}
  \label{eq:b.8}
  \varepsilon_{a b c} \, O_{b d} \, D_{d c} \stackrel{!}{=} 0 \ ,
\end{equation}
i.e. the matrix $\bbox{O}$ has to be chosen in such a way that for any given
diagonal matrix $\bbox{D}$ the matrix $\bbox{O} \cdot \bbox{D}$ is \emph{symmetric}. 
The only four solutions $\bbox{O} \in SO(3)$ of this problem turn out to be
\begin{eqnarray}
    \bbox{O}^{(1)} = \mbox{diag} (+1,-1,-1) & \qquad , \qquad &
   \bbox{O}^{(2)} = \mbox{diag} (-1,+1,-1) \ ,
  \nonumber \\
  \label{eq:b.9}
   \bbox{O}^{(3)} = \mbox{diag} (-1,-1,+1) & \qquad , \qquad &
   \bbox{O}^{(4)} = \mbox{diag} (+1,+1,+1) \ ,
\end{eqnarray}
\\
hence the ansatz (\ref{eq:b.5}) gives exactly \emph{four} further divergence-free triads
of Bianchi-type~IX homogeneous manifolds, 
\begin{equation}
  \label{eq:b.10}
  \vec{d}^{\; (\alpha)}_a = E_{b a} \, {{O}^{^{\mbox{\scriptsize{$(\alpha)$}}}} \!\!\!\!\!\!_{b c}} 
  \ \cdot \vec{\imath}_c \qquad ,
  \qquad \alpha \in \{ 1,2,3,4 \} \ .
\end{equation}

We now wish to compute the semiclassical saddle-point contributions to the vacuum state
(\ref{eq:4.31}), which correspond to the divergence-free triads 
$\vec{d}^{\; (\alpha)}_a , \alpha \in \{ 1,2,3,4 \}$. Therefore we first need the winding 
numbers $\hat w$ of these triads with respect to the Einstein-triad $\vec{g}_a$ of 
Bianchi-type~IX metrics. Since the Einstein-triad turns out to be given exactly by the 
invariant triad of the homogeneous 3-metric, $\, \vec{g}_a \equiv \vec{\imath}_a$, we have 
to calculate the Cartan-Maurer invariants (\ref{eq:4.0.2}) of the four rotation matrices
\begin{equation}
  \label{eq:b.11}
  \bbox{\Omega}^{(\alpha)} := \bbox{E}^{\mbox{\scriptsize{$T$}}} \!\! \cdot \bbox{O}^{(\alpha)}
  \qquad , \qquad \alpha \in \{ 1,2,3,4 \} \ .
\end{equation}
This can be done \emph{without} knowing the matrix $\bbox{E}$ in (\ref{eq:b.11})
explicitly, because the spatial derivatives in (\ref{eq:4.0.2}) may be substituted by
$\partial_j = \imath^{+}_{j a} \cdot \vec{\imath}^{\; +} \!\!\!\!_a$, and then be eliminated with
help of (\ref{eq:b.7}), yielding
\begin{equation}
  \label{eq:b.12}
  I ( \bbox{\Omega}^{(\alpha)} ) = - 8 \int d^3 x \, \varepsilon_{a b c} \, 
  \bbox{\imath}^{+} \!\!\!\!_{a} \wedge \bbox{\imath}^{+} \!\!\!\!_{b} \wedge
  \bbox{\imath}^{+} \!\!\!\!_{c} = -48 \, {\cal V} \ ,
\end{equation}
where ${\cal V} = 2 \, \pi^2$ is the dimensionless volume of the unit 3-sphere. Since the
constant $I_0$ in the definition (\ref{eq:4.0.3}) of the winding number has the numerical
value $I_0 = 96 \, \pi^2$ for manifolds with $S^3$-topology (cf.~\cite{Wein2}), it 
follows that the ``absolute'' winding numbers of the triads 
$\vec{d}^{\, (\alpha)}_a , \alpha \in \{ 1,2,3,4 \},$ are simply given by $\hat w = -1$. 

To proceed in the computation of the semiclassical 
saddle-point contributions (\ref{eq:4.31}), we
further have to  evaluate the functional $\phi$ defined in (\ref{eq:2.20a}) for the four
divergence-free triads $\vec{e}_a = \vec{d}^{\, (\alpha)}_a , \alpha \in \{ 1,2,3,4 \}$. 
Inserting the triads (\ref{eq:b.10}) into $\phi$ according to (\ref{eq:2.20a}), we first
recover the ``wormhole''-exponent of (\ref{eq:4.32.2}), 
if the spatial derivative $\partial_j$ acts on the
invariant triad $\vec{\imath}_a$. In addition, we obtain
a second term, which stems from the action of the derivative operator $\partial_j$
on the spatially nontrivial matrix $\bbox{E}$. This contribution can again be calculated
by reexpressing the spatial derivative in terms of the vector-fields 
$\vec{\imath}^{\; +} \!\!\!\!_a$, and making use of eqs.~(\ref{eq:b.7}). In case of the
divergence-free triad  $\vec{d}^{\, (4)}_a$, we obtain the explicit result 
\begin{equation}
  \label{eq:b.13}
  \Psi^{(4)}_{\rm vac} \, \stackrel{\mu \to \infty}{\propto}
  \Psi^{(0)}_{\rm vac} \cdot \exp \left [
  \pm \, \frac{ 4 {\cal V}}{\gamma  \hbar} \left (
  - \frac{6}{\Lambda} + a_1 \, a_2 + a_2 \, a_3 + a_3 \, a_1 \right ) \right ] \ ,
\end{equation}
with $\Psi^{(0)}_{\rm vac}$ given in (\ref{eq:4.32.2}). The saddle-point value
(\ref{eq:b.13}) is known as the ``no-boundary'' state from the homogeneous
Bianchi~IX model. Three further semiclassical saddle-point contributions to
the vacuum state (\ref{eq:4.31}), which correspond to the remaining 
divergence-free triads $\vec{d}^{\, (\alpha)}_a , \alpha \in \{ 1,2,3 \}$, are of the same 
form as $\Psi^{(4)}_{\rm vac}$ given in (\ref{eq:b.13}), but with two of the three
scale parameters ${a}_{b}$ replaced by their negatives. In the framework of the 
Bianchi~IX model, the corresponding  states were referred to
as ``asymmetric'' states. We conclude that all
five saddle-point values $\Psi^{(\alpha)}_{\rm vac} \, , \alpha \in \{ 0, \dots , 4 \},$ 
known for the 
homogeneous Bianchi~IX model can be recovered within the inhomogeneous approach
of the present paper by evaluating the state (\ref{eq:4.31}) for the five
topologically inequivalent divergence-free triads 
$\vec{d}^{\, (\alpha)}_a , \alpha \in \{ 0 , \dots , 4 \},$ of Bianchi-type~IX manifolds. Up to
a Gaussian prefactor, which always lies hidden in the proportionality signs of 
eqs.~(\ref{eq:4.32.2}), (\ref{eq:b.13}), the results are of the same form as
in \cite{Pat1,Pat3}. 
   
However, as we have shown in \cite{Pat3,Pat4}, the four semiclassical saddle-point
contributions $\Psi^{(\alpha)}_{\rm vac} \, , \alpha \in \{ 1,2,3,4 \},$ are restricted to
occur \emph{simultaneously} for symmetry reasons. This can also be seen within the
present, inhomogeneous approach, since the four divergence-free triads
$\vec{d}^{\, (\alpha)}_{a} , \alpha \in \{ 1,2,3,4 \},$ all have the same winding number,
and thus should enter into the value of the Chern-Simons state with the same 
topological right. We conclude that, in agreement with discussions of the
non-diagonal Bianchi~IX model, only \emph{two} 
independent values of the vacuum Chern-Simons state are found for 
Bianchi-type~IX manifolds.


\section{A non-flat 4-metric generated by the vacuum state}
\label{appC}

We now want to give special solutions of the vacuum
evolution equations (\ref{eq:4.40.0}), such that the
associated semiclassical 4-geometries satisfy neither the
``no-bondary'' condition proposed by Hartle and Hawking \cite{Haw1,Haw2,Haw3},
nor the condition of asymptotical flatness in the limit of large 
scale parameters $a_{\rm cos}$.\footnote{Here we assume the vacuum limit $\kappa \to 0$
to be realized by considering a sufficiently small value for the cosmological constant 
$\Lambda$. Then it will be possible to take the cosmological scale parameter $a_{\rm cos}$ 
arbitrarily large at the same time, cf.~eq.~(\ref{eq:4.1.4}).} 
Let us therefore consider the class of 3-metrics
\begin{equation}
\bbox{h} = \vec{\imath}_a \otimes \vec{\imath}_a \ ,
\label{eq:c1}
\end{equation}
where the triad vector-fields $\vec{\imath}_a = {\imath}^{\, i} \,\!_a \, \partial_i$ are given by
\begin{equation}
 \vec{\imath}_1 = \frac{1}{a_1} \, \partial_1 \ \ ,\ \ 
 \vec{\imath}_2 = \frac{1}{a_2} \, \partial_2 \ \ ,\ \
 \vec{\imath}_3 = \frac{1}{a_3} \left ( \partial_3 + x^2 \, \partial_1 + x^1 \, \partial_2 \right ) \ .
\label{eq:c2}
\end{equation}
The scale-parameters ${a}_{b}$ in (\ref{eq:c2}) are assumed to be spatially constant,
and the triad $\vec{\imath}_a$ is taken to be positive-oriented.
Then the structure matrix $\bbox{m}$ introduced in
(\ref{eq:4.16a}) takes the spatially constant form
\begin{equation}
\bbox{m} = \mbox{diag} \left [ \, \frac{a_1}{a_2 a_3} \, , \,  -\frac{a_2}{a_3 a_1} \, , \, 0 
\, \right ] \ ,
\label{eq:c3}
\end{equation}
i.e. the triad $\vec{\imath}_a$ is the invariant triad of a
spatially homogeneous 3-manifold, which can be classified to be of Bianchi-type VI$_{-1}$.
Since the structure matrix $\bbox{m}$ according to (\ref{eq:c3}) is symmetric, it follows
directly from eq.~(\ref{eq:4.32}) that the invariant triad $\vec{\imath}_a$ is divergence-free.
The Killing-vectors of the 3-metric (\ref{eq:c1}) must commute with the $\vec{\imath}_a$
and are given by
\begin{equation}
 \vec{\xi}_1 = \cosh x^3 \, \partial_1 + \sinh x^3 \,  \partial_2 \quad , \quad
 \vec{\xi}_2 = \sinh x^3  \, \partial_1 + \cosh x^3 \, \partial_2 \quad , \quad
 \vec{\xi}_3 = \partial_3 \ .
\label{eq:c5}
\end{equation}
They may be used to compactify the 3-manifold ${\cal M}_3$ with the metric
(\ref{eq:c1}) in the three $\vec{\xi}_a$-directions, giving rise to a manifold with the
nontrivial topology $S^1 \times T^2$. The compactified 3-manifold will then have a 
finite volume $V = {\cal V} \, a_1 \, a_2 \, a_3$, where the value of ${\cal V} > 0$ 
depends on the 
particular choice of the compactification.

We are now interested in the semiclassical 4-geometries being generated by
the evolution equations (\ref{eq:4.40.0}) in case of the divergence-free triad 
$\vec{d}_a = \vec{\imath}_a$. If we allow for an arbitrary
lapse-function $N$, they read
\begin{equation}
\frac{d}{d\tau} \, \tilde{\imath}^{\, i} \,\!_{a} = \pm N \, \tilde{\varepsilon}^{i j k} \, \partial_j \, 
\imath_{k a} \ .
\label{eq:c5a}
\end{equation}
For the three-metric (\ref{eq:c1}) under study, eqs.~(\ref{eq:c5a}) take
the form
\begin{equation}
 \frac{d}{d\tau} \, \sigma_1 = \mp N \, \sqrt{\frac{\sigma_2 \, \sigma_3}{\sigma_1}} \quad , \quad
 \frac{d}{d\tau} \, \sigma_2 = \pm N \, \sqrt{\frac{\sigma_3 \, \sigma_1}{\sigma_2}}\quad , \quad
 \frac{d}{d\tau} \, \sigma_3 = 0 \ ,
\label{eq:c6}
\end{equation}
where we have introduced the new variables
\begin{equation}
 \sigma_1 := a_2 \, a_3 \quad , \quad
 \sigma_2 := a_3 \, a_1 \quad , \quad
 \sigma_3 := a_1 \, a_2 \ .
\label{eq:c7}
\end{equation}
Choosing the lapse-function $N$ as
\begin{equation}
 N = \mp {\textstyle \frac{1}{2}} \, \left ( \sigma_1 \, \sigma_2 \, \sigma_3 \right )^{-1/2} \ ,
\label{eq:c8}
\end{equation}
the set of eqs.~(\ref{eq:c6}) is easily integrated and has the general
solution
\begin{equation}
\sigma_1 (\tau) = \sqrt{\tau_0 + \tau} \quad , \quad
\sigma_2 (\tau) = \sqrt{\tau_0 - \tau} \quad , \quad
\sigma_3 (\tau) \equiv \sigma_3 = \mbox{const.} \quad ; \quad
| \tau | < \tau_0 \ .
\label{eq:c9}
\end{equation}
Here we have chosen $\tau=0$ such that $\sigma_1 (0) = \sigma_2 (0)$, so only
two integration constants $\tau_0 > 0$ and $\sigma_3 > 0$ remain in (\ref{eq:c9}).

In order to prove that the 4-geometry according to (\ref{eq:c9}) is non-flat,
it is \emph{not} sensible to compute the 4-dimensional Ricci- or Einstein-tensor,
since these quantities vanish identically  by construction, so we will consider the
nontrivial components  
$^4 \! {R^{0 i}}_{0 j}$ of the
4-dimensional Riemann-tensor instead. For a vanishing shift-vector $N^i=0$, 
they are given by
\begin{equation}
 ^4 \! {R^{0 i}}_{0 j} = - \frac{1}{N} \frac{d}{d t} \, {K^i}_{j} + {K^i}_{k} \, {K^k}_{j} \ ,
\label{eq:c10}
\end{equation}
with ${K^i}_{j}$ being the usual extrinsic curvature tensor. 
With help of the triad (\ref{eq:c2}), we may convert the spatial indices of
$^4 \! {R^{0 i}}_{0 j}$ into internal indices $a$, $b$, to obtain
\begin{equation}
 {\cal R}_{a b} := {\imath}_{i a} \, {\imath}^{\, j} \,\!_b \, ^4 \! {R^{0 i}}_{0 j} \ .
\label{eq:c12}
\end{equation}
For the metric (\ref{eq:c1}), $( {\cal R}_{a b} )$ is a diagonal matrix with
\begin{equation}
 {\cal R}_{3 3} =
 -\frac{1}{N \, a_3} \, \frac{d}{d\tau}
 \left ( \frac{1}{N} \, \frac{d a_3}{d \tau} \right ) \ ,
\label{eq:c13}
\end{equation}
\\
and analogous expressions for ${\cal R}_{1 1}$, ${\cal R}_{2 2}$. Making use of
the evolution eqs.~(\ref{eq:c6}), we can eliminate the $\tau$-derivatives
in (\ref{eq:c13}) to arrive at
\begin{equation}
{\cal R}_{3 3} = \frac{\sigma_1 \, \sigma_2 \, \sigma_3}{2}
\left ( \frac{1}{\sigma_1^2} + \frac{1}{\sigma_2^2} \right )^2 \ ,
\label{eq:c14}
\end{equation}
and inserting the general solution (\ref{eq:c9}) into (\ref{eq:c14}), we find
\begin{equation}
 {\cal R}_{3 3} = \sigma_3 \, \tau_0^2 \left ( \tau_0^2 - \tau^2 \right )^{-3/2} > 0 \ .
\label{eq:c15}
\end{equation}

Thus we have found a component of the Riemann-tensor, which is nonzero for all
times $\tau$, $| \tau |< \tau_0$, so the semiclassical 4-geometries obtained
by evolving the initial 3-geometries (\ref{eq:c1}) are \emph{nowhere} flat. Moreover, the 
semiclassical 4-geometries do \emph{not} satisfy the ``no-bondary'' condition:
While the cosmological scale parameter
\begin{equation}
  \label{eq:c16}
  a_{\rm cos} = {\cal V}^{1/3} \left ( \sigma_1 \, \sigma_2 \, \sigma_3 \right )^{1/6} =
 {\cal V}^{1/3} \, \sigma_3^{1/6} \left (\tau_0^2 - \tau^2 \right )^ {1/12}  
\end{equation}
vanishes only at the timelike borders $| \tau | \to \tau_0$ of the semiclassical
space-time manifolds, 
the corresponding curvature components 
${\cal R}_{3 3}$ at the same time are tending to $+ \infty$. Consequently, the 
semiclassical 4-manifolds are \emph{not} regular or
compact for vanishing scale parameters $a_{\rm cos}$.

\end{appendix}


\end{document}